# Evaluating Transit Accessibility to Education and Effects of Operational Delays in Japanese Regional Cities: A Case Study of Matsumoto City

Itsuki Sato[1*], Kiyoshi Takami[1] and Giancarlos Parady[1]

[1] Department of Urban Engineering, The University of Tokyo,
7-3-1 Hongo, Bunkyo-ku, Tokyo, Japan.
*Corresponding author. E-mail: mailto:itsuki.rail@gmail.com;
Contributing authors:
takami@ut.t.u-tokyo.ac.jp; gtroncoso@ut.t.u-tokyo.ac.jp

**Abstract**

Realistic assessments of school commuting accessibility in areas with infrequent public transport services require accounting for operational delays; however, the impact of these delays has not been sufficiently examined. This study evaluates high-school accessibility in Matsumoto City, a regional city in Japan, using GTFS data representing both scheduled timetables and actual operating conditions. Accessibility levels are assessed under scheduled operations, while the effects of delays are examined through a comparative analysis based on actual delay measurements over a five-day workweek. Furthermore, a sensitivity analysis of travel-time thresholds was conducted. Results show that, when walking, cycling to stations, and public transport use are allowed, 78% of children under 15 can reach at least one high school within a 90-minute round trip, and 67% within a 60-minute round trip. Extending the threshold to 120 minutes enables access to nearly all schools in the city center, but the overall proportion increases only marginally to 81%. Delay impacts are particularly pronounced along bus routes connecting the central station with suburban areas, while in some areas, delays generate idiosyncratic events, where irregular transfers and reduced waiting times result in improved accessibility. Results underscore the need for both short-term measures—such as adjusting school start times, prioritizing buses, and introducing dedicated school routes—and long-term strategies—such as incorporating public transport accessibility into school consolidation decisions—to guarantee fair access to education opportunities without relying on private vehicles.

# 1 Introduction

## 1.1 Background

Accessibility can be defined as the potential of opportunities for interaction (Hansen 1959) or the freedom of individuals to decide whether to participate in different activities (Burns 1979). As such, it is a critical index for evaluating the service quality of transportation infrastructures that support daily life. A substantial number of accessibility studies have focused on public transport as a means of enhancing the quality of life for individuals who are unable or unwilling to travel by car (Hasegawa et al. 2022; Carney et al. 2022; Goliszek 2022; Torres and McArthur 2024).

Geurs and van Wee (2004) identify four types of accessibility measures: *infrastructure-based*, *location-based*, *person-based*, and *utility-based*. In this study, we adopt location-based measures since these offer the most appropriate balance of conceptual rigor and practical use. Location-based measures fall in the category of integral accessibility, which evaluates access to multiple activity opportunities rather than only transit stops or a single destination (Malekzadeh and Chung 2020). In contrast, Japanese transport planning often relies on buffer-based measures that classify areas by distance to transit stops (i.e. *system accessibility* as defined in Malekzadeh and Chung 2020). However, because accessibility fundamentally concerns reaching activity opportunities, proximity to transit alone provides an incomplete assessment.

Furthermore, service punctuality is an important consideration in accessibility analyses. Delays can occur for a variety of reasons and affect both private car and public transport travel. In particular, in regional cities with infrequent bus services, the ability to catch a bus can significantly alter arrival times at destinations, making such delays a potential threat to users' convenience (Nichols et al. 2024). This suggests that evaluations based on predetermined schedules may not account for this uncertainty, resulting in inaccurate and overly optimistic representations of actual conditions (Wessel 2019).

Fortunately, recent advances in data collection and provision, including GTFS Realtime and other real-time bus location data, have made it possible to analyze accessibility based on actual operational conditions. There has also been an emergence of fast travel-time calculation engines with scheduled services using data inputs such as OpenStreetMap and GTFS (Conway et al. 2018; Pereira et al. 2021; Higgins et al. 2022; Torres and McArthur 2024). When combined with delay data, it is now possible to calculate accessibility more closely aligned with current conditions than ever before. In this regard, previous studies have demonstrated the impact of bus delays on accessibility analysis by comparing accessibility between scheduled and delayed services (Liu et al. 2023; Nichols et al. 2024). In Japan, a number of studies have examined public transport accessibility, but most focus on urban areas and regional core cities (Tanimoto et al. 2009; Kiyohara and Yanagihara 2020; Hasegawa et al. 2022). For example, Tanimoto et al. (2009) considered constraints from limited service frequencies, while Kiyohara and Yanagihara (2020) and Hasegawa et al. (2022) incorporated average waiting time. However, most studies have overlooked factors such as bus delays, real-time operational data, round-trip travel, and access to educational opportunities, leaving key aspects of public transport accessibility underexplored. Especially, round-trip-based accessibility analysis remains limited, despite the substantial differences in service frequencies between outbound and return trips. While this difference may have only a minor effect in cities with high-frequency services, it can significantly influence accessibility levels in areas with infrequent public transport services. Therefore, incorporating round-trip accessibility is important for achieving a more accurate assessment of public transport accessibility.

These gaps are increasingly important in light of recent policy developments. Since April 2024, the Ministry of Health, Labor and Welfare has strengthened labor standards for bus drivers, reducing service frequency (Suwa 2024). At the same time, the Ministry of Land, Infrastructure, Transport and Tourism (MLIT) has promoted the redesign of local public transport networks, emphasizing efficiency, sustainability, and multimodal coordination. These initiatives are institutionalized through Regional Public Transport Plans, and MLIT provides guidance, consulting, and regulatory support, including joint route operations and vertical separation between infrastructure and operations. As of November 2025, more than 1,200 Regional Public Transport Plans have already been formulated (MLIT 2025).

In local cities with low-frequency services, delays can strongly affect arrival times and accessibility. Evaluating these effects is thus crucial, particularly for non-drivers such as the elderly and students enrolling in high schools, and to inform the design of restructured public transport networks (Yanagisawa et al. 2022).

## 1.2 Research objectives

This study aims to clarify whether public transport services in a Japanese regional city adequately meet the needs of citizens who cannot drive, particularly high schools for students, by explicitly considering actual operational conditions in accessibility analysis. Specifically, we:

(1) Investigate public transport accessibility to high schools based on a static timetable at both high-resolution (250m) regular grid and administrative district levels.

(2) Evaluate accessibility that reflects actual operational conditions through weekdays in a specific week, including delays, and quantify the effect of delays on accessibility.
(3) Discuss the implications of the results from (1) and (2) for students and households reliant on public transport.

While Japan has a well-developed public transport system, accessibility analyses that account for operational conditions remain scarce, particularly in regional cities where services are less frequent and public transport is struggling to survive. This study aims to quantify the loss of activity opportunities, particularly for students, due to delays, and highlights the need for quantitative analyses that explicitly consider operational conditions.

The structure of this paper is as follows. Section 2 reviews (i) the literature on accessibility analysis in Japan and abroad that considers delays, and (ii) policy documents and studies focused on educational opportunities. Section 3 firstly describes the current status of bus services in the case study area; then it introduces the methodology of travel-time and index calculation, containing a brief summary of the necessary data sources. Section 4 quantifies accessible educational opportunities to high schools under scheduled public transport services at grid and district level. Section 5 evaluates accessibility to high schools under actual operational conditions and the gains or losses due to delays. Finally, Section 6 discusses the findings and their implications for educational policy and highlights areas for future research.

# 2 Literature review

## 2.1 Travel-time calculation for accessibility analysis

A key question in accessibility analysis is how to calculate travel time to each destination. One traditional approach is to calculate the sum of the in-vehicle travel time and the expected waiting time (Hilman and Pool 1997), with waiting times typically estimated as half the headway of a transit service. In Japan, this method was introduced under the name of T-index, to evaluate accessibility levels at each point in cities (National Institute for Land and Infrastructure Management 2014). Kiyohara and Yanagihara (2020) adopted the T-index to illustrate changes in accessibility to lifestyle-supporting facilities (e.g., supermarkets, hospitals, and clinics) before and after the reorganization of transportation services in Maibara City, a regional city in Japan. This index has also been used in analyses of public transport accessibility with subway stations as destinations (Goto and Takahashi 2023). Some local municipalities have also utilized this method to evaluate public transport convenience (e.g., Kasugai City 2021; Otsuki City 2023; Sakura City 2024).

While this method might be suitable for areas with frequent and convenient services in which passengers do not need to concern themselves with timetables, such as the Greater Tokyo Area, its practicality is considered insufficient in regional cities with limited public transport. This is because citizens in areas with less frequent services are more likely to plan their trips according to the timetable than to leave their origin point at random times.

Another approach is to essentially ignore the waiting time before the first boarding and instead impose a certain waiting time penalty when using public transport (Department for Transport 2019). This method, however, does not accurately account for waiting times, which might lead to overestimating the convenience of public transport.

The third method involves setting a time window for departure, performing route calculations using open-data sources, and determining representative travel times. Fortunately, access to public transport data for routing engines is improving as GTFS format grows in popularity. This makes schedule-based accessibility evaluations easier. Using GTFS data, one can calculate average cumulative opportunities across a certain time window (Owen and Levinson 2014; Farber and Fu 2017). Conway et al. (2018) pointed out the problems associated with evaluation accessibility using averages and proposed a new method using median or percentile travel times in time windows. For a more comprehensive review of the accessibility literature, the reader is referred to Malekzadeh and Chung (2020).

## 2.2 Delay consideration on accessibility studies

To incorporate actual operating conditions, such as delays, into accessibility analyses, tracking technology, such as GPS, is necessary (Braga et al. 2023). Research in this area has progressed in recent years with the development of bus location systems and GTFS Realtime. Wessel and Farber (2019) collected five days of vehicle GPS data to demonstrate variability in public transport travel time. Additionally, schedule-based catchment areas have been compared with GTFS-based catchment areas that reflect actual operational conditions. Hybrid isochrones incorporate actual operational times into schedule-based routes (Liu et al. 2023). Another study compares schedule-based and delay database approaches to evaluate the accessibility of employment via public transport (Nichols et al. 2024).

However, research in Japan has mostly focused on the causes of public transport delays and users' perceptions of delays. For example, Takeshita et al. (2009) examined how concerns about service punctuality influenced travel behavior following the discontinuation of a new transit system (a rubber-tired Automated Guideway Transit) and compared accessibility before and after the system closure. While they calculated accessibility for survey respondents, they did not explicitly incorporate actual delays into their travel-time analysis and, therefore, did not quantify how decreased punctuality affected accessibility. More recently, some studies have used GTFS data to visualize spatiotemporal delay patterns (Kasahara et al. 2022) or to estimate delay conditions using GTFS Realtime data (Tajima et al. 2021). These studies, however, have focused on delay visualization rather than integrating delays into accessibility analysis. As such, conducting accessibility analyses that explicitly consider delay conditions remains an important research gap in the Japanese context.

## 2.3 Accessibility analysis in educational context

Various literature has emphasized the importance of educational opportunities in accessibility analysis. In England, Social Exclusion Unit (2003) discusses the relationship between transport services and land use from the perspective of social exclusion. The unit reports that loss of learning opportunities, such as being unable to attend one's preferred school, is also a matter of consideration. Furthermore, to address social exclusion issues, Department for Transport (2004) incorporated school commuting into accessibility planning considerations, and since then, accessibility to educational opportunities ranging from primary school to sixth form colleges has remained a focus of attention (Department for Transport 2019).

In the United States, a practical guide for accessibility evaluation by transportation agencies illustrates schools as one of several important destinations (National Academies of Sciences, Engineering, and Medicine 2022). Accessibility analyses have also focused on access to high schools and discussed regional disparities in other countries such as Brazil, Poland, and Korea (Moreno-Monroy et al. 2021; Rosik et al. 2021; Kim 2025). These analyses underscore the importance of using educational opportunities as a case study in accessibility research.

In Japan, particularly in regional cities, public transport services have deteriorated severely due to decrease in ridership and driver shortages (Suwa 2024). Conversely, while some studies consider accessibility to educational facilities as part of lifestyle-supporting facilities (Okuda et al. 2018; Yanagisawa et al. 2022), few explicitly address disparities in accessibility between regions quantitatively.

Additionally, school mergers and closures have become commonplace due to the declining population. This makes ensuring opportunities and the appropriate allocation of public transport a policy issue of the topmost importance (Nishimura et al. 2014). Therefore, it is important to measure commuting opportunities to high schools where students can access multiple educational facilities.

## 2.4 Contribution of this research

In summary, the reviewed literature highlights several key points. First, traditional travel-time calculation methods, such as the T-index, are widely used in Japan to evaluate accessibility, but they often rely on scheduled or average travel times and do not account for actual delays, which is particular in regional cities with infrequent services. Second, while recent advances using GTFS and GTFS Realtime data allow visualization of spatiotemporal delay patterns, these approaches have rarely been applied to quantify accessibility under realistic operating conditions. Third, educational facilities, particularly high schools, are important destinations in accessibility research, yet Japanese studies have seldom assessed disparities in access to these facilities quantitatively, especially in the context of declining public transport services and school closures.

Building on these gaps, this study explicitly incorporates operational conditions, including delays, into accessibility analysis to high schools in a regional city. By doing so, it contributes to the literature by (i) providing a quantitative assessment of the impact of delays on accessibility for non-drivers, (ii) applying location-based accessibility measures in combination with real operational data, and (iii) highlighting policy-relevant insights for local transport planning and educational access in regional Japanese cities.

# 3 Methodology and data

## 3.1 Overview of the case study area

This study uses Matsumoto City in Nagano Prefecture, Japan as a case study. This regional city has a population of approximately 230,000 and a city planning area, which refers to areas designated by a prefecture where the *City Planning Act* and *Building Standards Act* can limit and control development and building activities, of 301.91 km$^2$. The rules about transportation and land use are also well developed in these city planning areas.

Matsumoto Station, located at the center of the planning area, functions as the main hub of the city. Figure 1 shows the distribution of the population aged 14 and under, which is the focus of accessibility analysis. While a high concentration of children is observed in the periphery surrounding Matsumoto Station, notable clusters are also found in the southern and western parts of the city. The city planning area is divided into 11 regions and 32 districts, as shown in Figure 2. These administrative units are used as the aggregation units in this study.

In Matsumoto City, as in many other regional cities in Japan, bus routes have been reduced in response to declining ridership. However, based on the recognition of local public transport as a form of "social infrastructure," the city has taken an active role in reorganizing its bus network. Specifically, bus services that were previously operated by private companies have transitioned to a publicly owned, privately operated system, alongside the promotion of a comprehensive public transport plan that includes a large-scale reorganization of the bus route network. Previously oriented toward long-distance routes, the network has been reorganized into a more flexible, district-focused system, enhancing access to facilities designated under the *City Planning Act* and the *Location Optimization Plan* (Matsumoto City, Yamagata Village and Asahi Village 2023).

Figure 3 shows the number of bus services on each route (see Appendices 1 and 2 for the detailed bus frequency by route). Bus routes are broadly classified into two types: Regular Routes, which operate medium- and large-sized vehicles connecting the city center with surrounding areas, and Local Routes, which use smaller vehicles to provide intra-community circulation similar to community buses. Service frequencies are highest around Matsumoto Station, with some sections offering up to 100 trips per day in both directions, whereas suburban areas typically have 10–20 trips per day. Some Local Routes operate on alternate days or follow different routes depending on the day, as indicated by the blue lines in Figure 3.

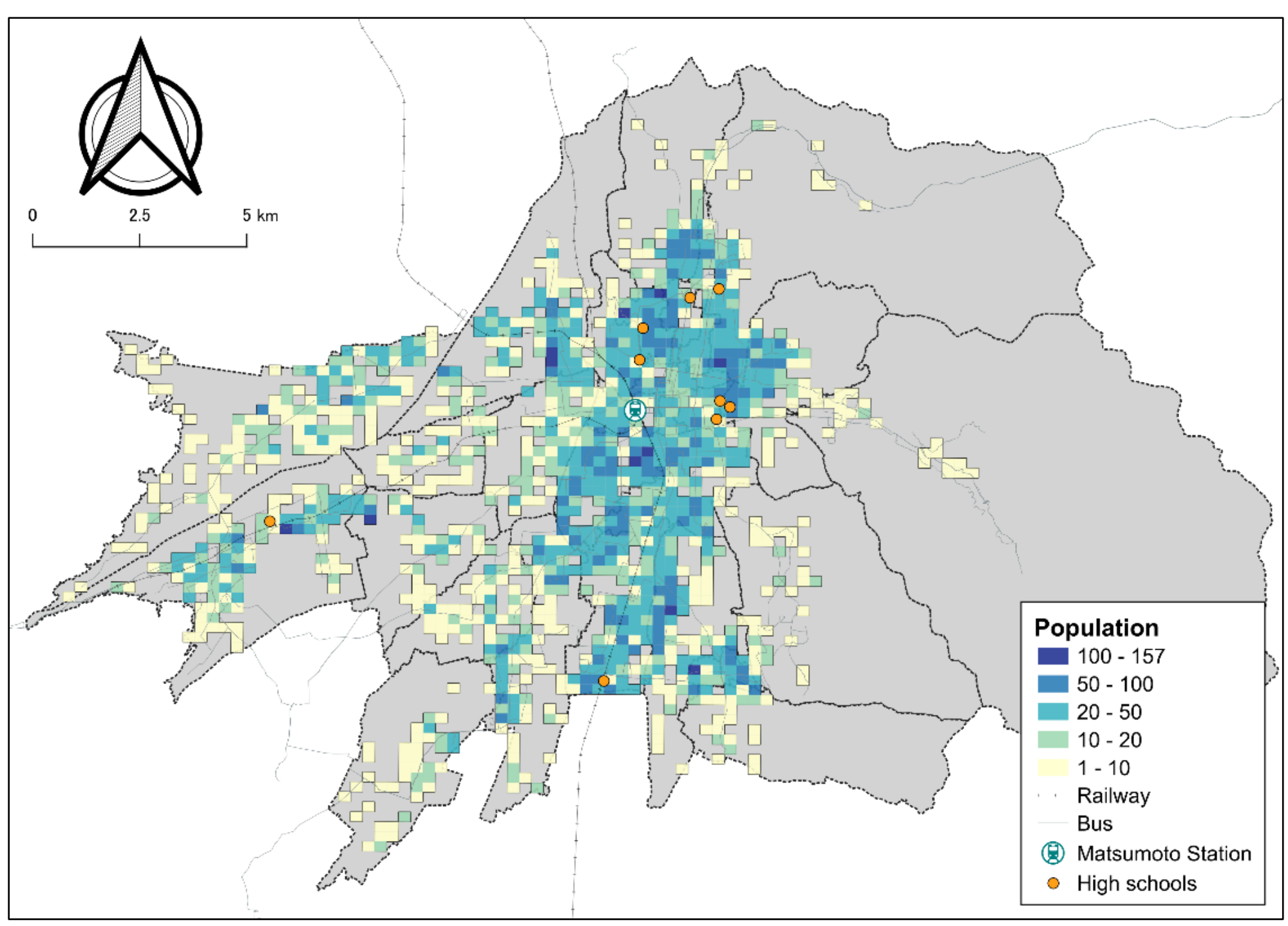

**Fig. 1** U15 population distribution in the city planning area of Matsumoto City

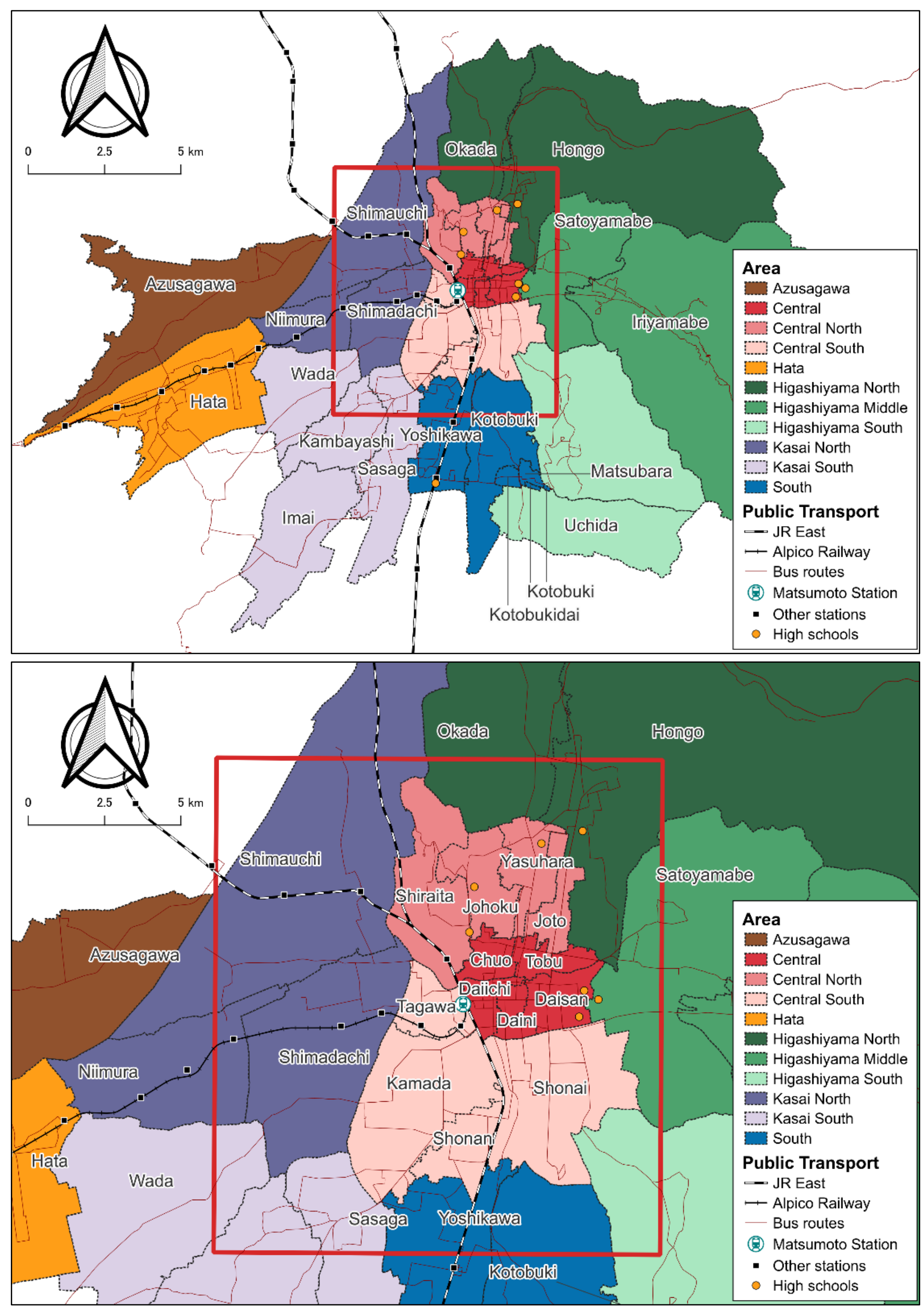

**Fig. 2** Names of districts and public transport network routes in Matsumoto City

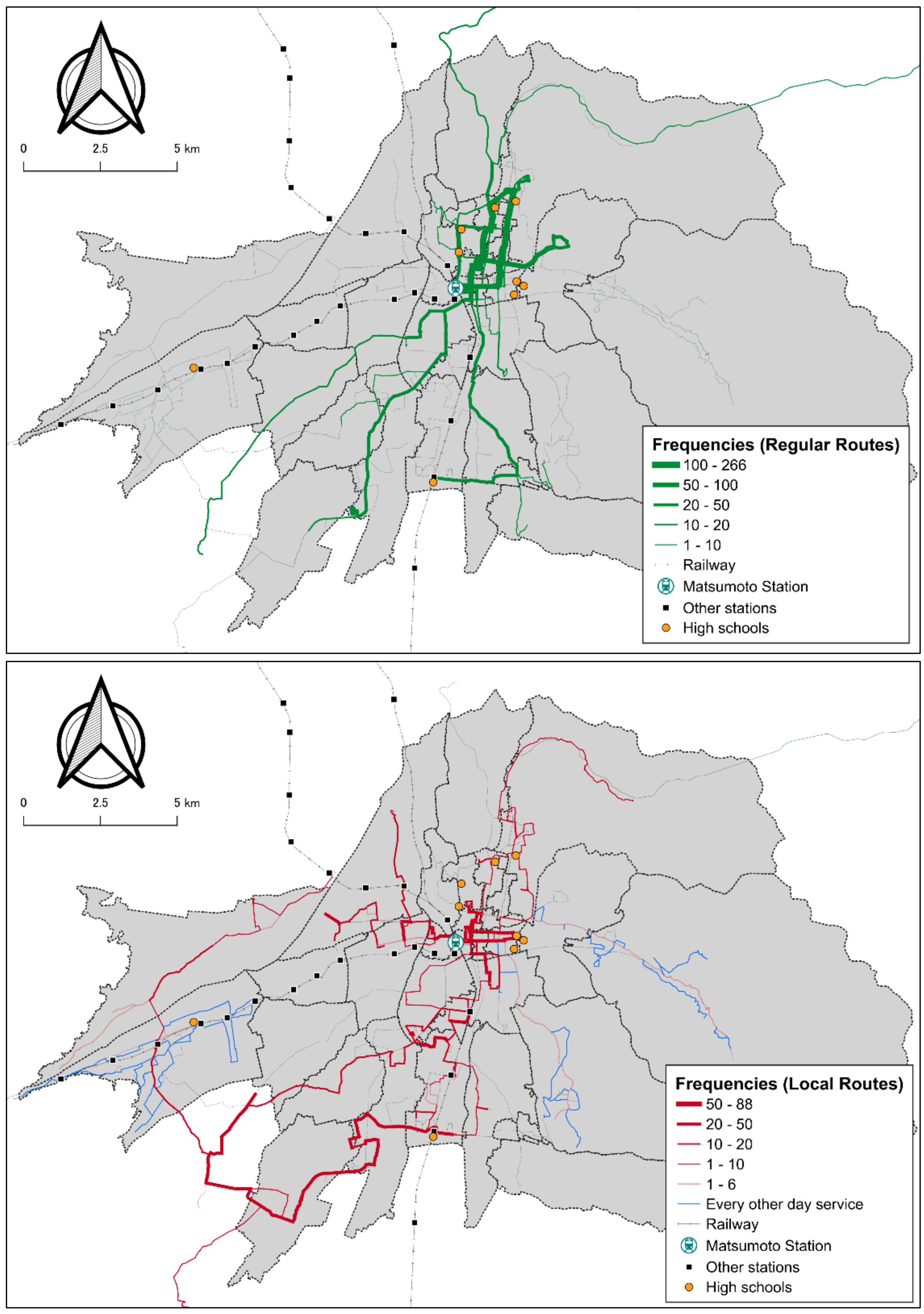

**Fig. 3** Comparison of Bus Frequencies: Regular Routes vs. Local Routes

This study conducts an accessibility analysis of educational facilities, with a particular focus on access to general academic high schools. In Matsumoto City, school districts are designated for public elementary and junior high schools, with the exception of private schools (Matsumoto City 2023). In other words, students' places of residence determine which elementary or junior high school they attend, leaving little room for households with children to benefit from improved access to multiple elementary or junior high schools.

By contrast, the number of high schools is smaller than that of elementary and junior high schools, and no school district system applies. As a result, students are able to choose any high school within the city. Moreover, high schools operate under diverse educational policies and offer distinctive curricula and school events. Consequently, securing access to multiple high schools represents an important advantage for households with children and contributes to the overall attractiveness of residential areas.

In Nagano Prefecture, based on the First High School Reorganization Plan, a series of consolidations and closures of high schools—including the establishment of integrated lower and upper secondary schools—were implemented between 2008 and 2021. During this process, accessibility was considered as one of the planning factors, while concerns were also raised by elementary and junior high school stakeholders regarding potential difficulties in students' commuting. Given this background, examining accessibility to high schools in regional cities is both important and meaningful.

In parallel, a web-based survey on bus usage awareness conducted across all age groups in Matsumoto City revealed that 14% of the respondents cited bus punctuality as the second most important reason for inconvenience (Matsumoto City 2021). These findings highlight the relevance of evaluating accessibility under realistic operational conditions, including delays, to ensure that public transport adequately meets the needs of those who cannot drive.

## 3.2 Overall flow of the study

Figure 4 illustrates the overall workflow of this study. When information on origins and destinations (OD) and transportation networks is input, a program created using the r5r package (Conway et al. 2018; Pereira et al. 2021) in the R language estimates travel-time matrices and opportunity indices. For each day considered in the analysis, these calculations are performed under two conditions: (i) static conditions based on GTFS data derived from schedules and (ii) dynamic conditions based on GTFS data reflecting actual arrival and departure times (including delays) at each stop. Based on these travel times, the indices described in Subsection 3.5 are calculated, and differential calculations or index comparisons are performed.

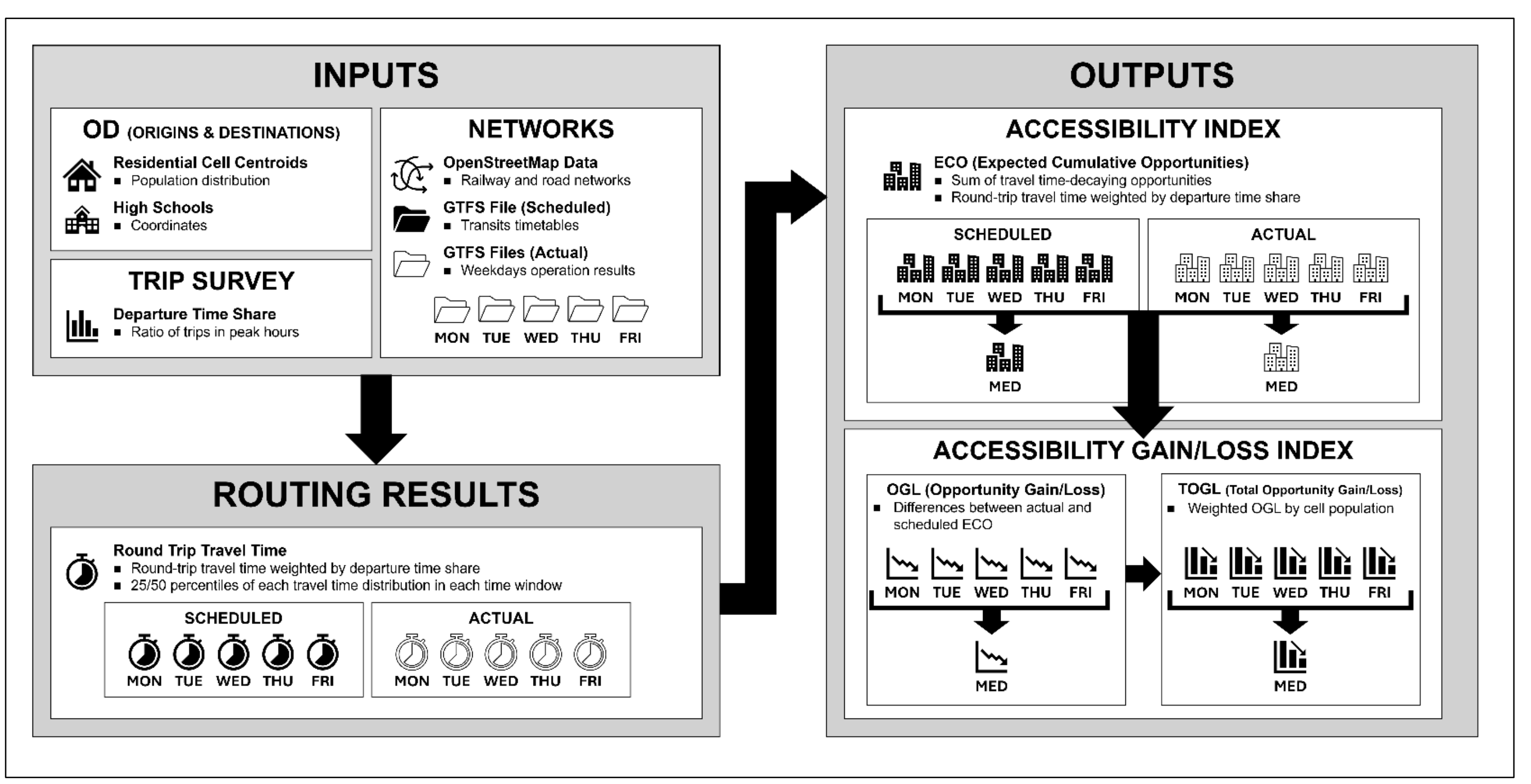

**Fig. 4** Study Workflow

## 3.3 Collection and generation of data

### 3.3.1 Data on origins and destinations

This study analyzes accessibility between residences and high schools, both of which are regarded as ODs. Residence points were determined by extracting a 250m-resolution grid with a non-zero population aged 14 and under from the *Population Census 2020* (Statistics Bureau of Japan 2021) and setting the centroid of each grid as the representative point. This study focuses on the 11 areas and 32 districts within the city planning area in Matsumoto City, shown in Figure 2. Each cell is tagged with the name of the region and district, which is used to calculate the population percentile values described below.

High school coordinates target the high schools within the city that offer general education programs. A general education program is a type of high school in Japan that provides general secondary education. There are also vocational programs, such as industrial and commercial programs, as well as a program that focuses on preparing for university entrance exams. General high schools are open to all students regardless of their interests and abilities, which is appropriate for measuring accessibility to educational opportunities.

### 3.3.2 Data on transportation network and operation

The public transport network of Matsumoto City consists of three railway lines and the "Gurutto Matsumoto Bus," an integrated bus system operated through a public-private partnership. GTFS data related to these services are required for analysis.

The bus GTFS data on static timetables was obtained from the city's open data portal (https://gtfs-data.jp/). The railway GTFS data was generated by the authors for the regular trains, all of which have stops within the city planning area. The static timetable information from the March 2025 revision was used as a reference and no delays were assumed.

Regarding the data on the buses' actual operational status, we extracted the data from the "Shinshu Navi" web bus location system service (https://qng.jcld.jp/NaganoRouteFinderNext/) operated by the prefectural government, which provides sightseeing guides and bus information of the city. The system acquires vehicle stop data and notifies users of updates to the icon position every five seconds via an *XMLHttpRequest* object. Data were collected every 30 seconds on vehicle ID, names of previous/next stops, departure time of the previous stop, delay time, etc. on weekdays in a specific week (December 22nd-26th 2025) and generated the GTFS data which reflected actual arrival and departure times. The measurement dates were set for these days to avoid the effects of holiday traffic congestion. In addition, the semester at high schools in Matsumoto City continued until Friday, December 26 before winter break, indicating that the survey period coincided with a time when students were actively commuting to school.

As shown in Figure 5, it was essential to respond to irregular situations in generating GTFS data as follows. When the actual operational data was missing due to communication lags, the departure time recorded immediately before the departure time was added to the predetermined duration to reach the stop on the schedule. However, the vehicle's arrival at a stop may occur earlier than anticipated due to the vehicle passing a stop. In such circumstances, the incorporation of the stipulated duration may result in the sequence of departure times between stops being inverted. For simplicity, the departure time was set to be the same as that of the subsequent stop in such cases. Missing data replacement works as this flow shows to avoid a reversal in departure times between ordered bus stops.

The distributions of median delay times for each bus route, derived from collected and processed delay data, are shown in Figures 6 and 7. The values shown in both figures represent the medians of the average delays calculated for each service over the five-day period. Figure 6 illustrates the delay distribution for Regular Routes, where the green points represent the median delay, and the error bars indicate the observed range (from minimum to maximum). A number of routes experience delays of 5–10 minutes, which may be exacerbated by the large number of passengers typically carried on medium- and large-sized vehicles.

Figure 7 presents the delay distribution for Local Routes, with red points indicating the median value. The Town Sneaker lines, which circulate within the central urban area, show slightly larger delays, but these generally remain around 5 minutes. Although the General Administration Liner is categorized as one of the Local Routes, it operates with large vehicles and serves a relatively high number of passengers, resulting in longer delays. Overall, most local routes experience delays of 0–5 minutes.

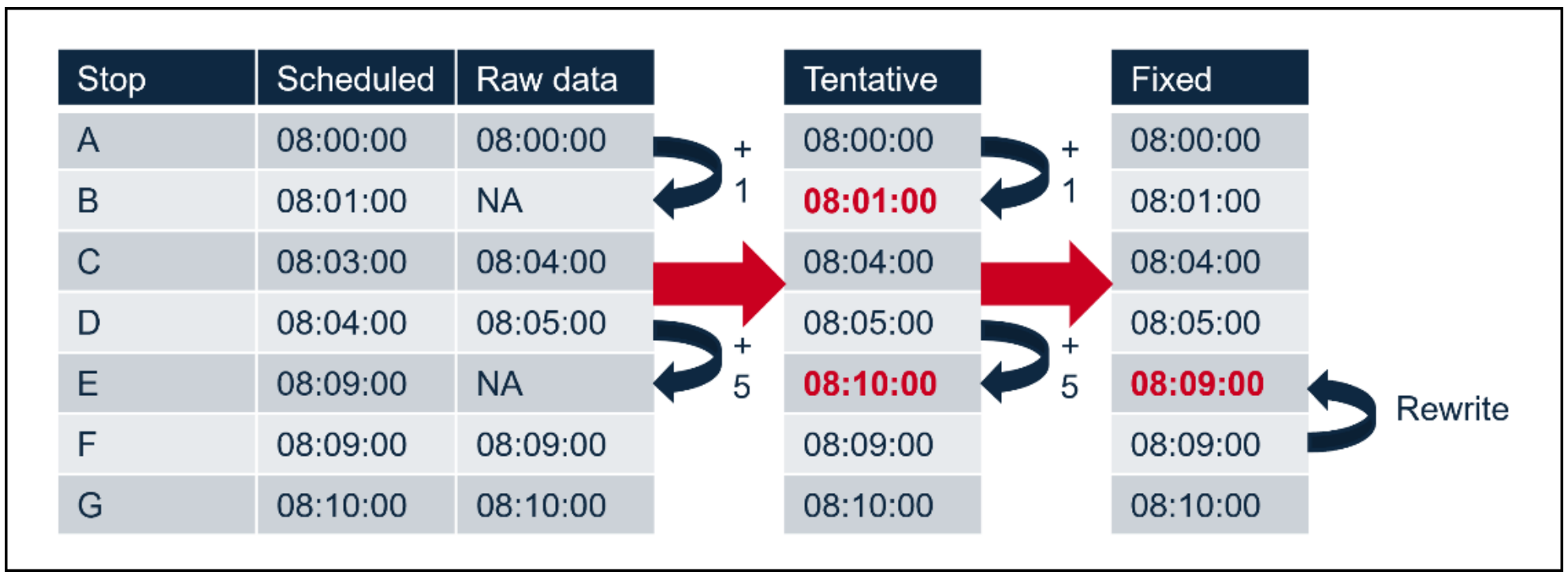

**Fig. 5** Example of delay implementation into GTFS data (stop_times.txt) and correction of irregularities produced during data extraction

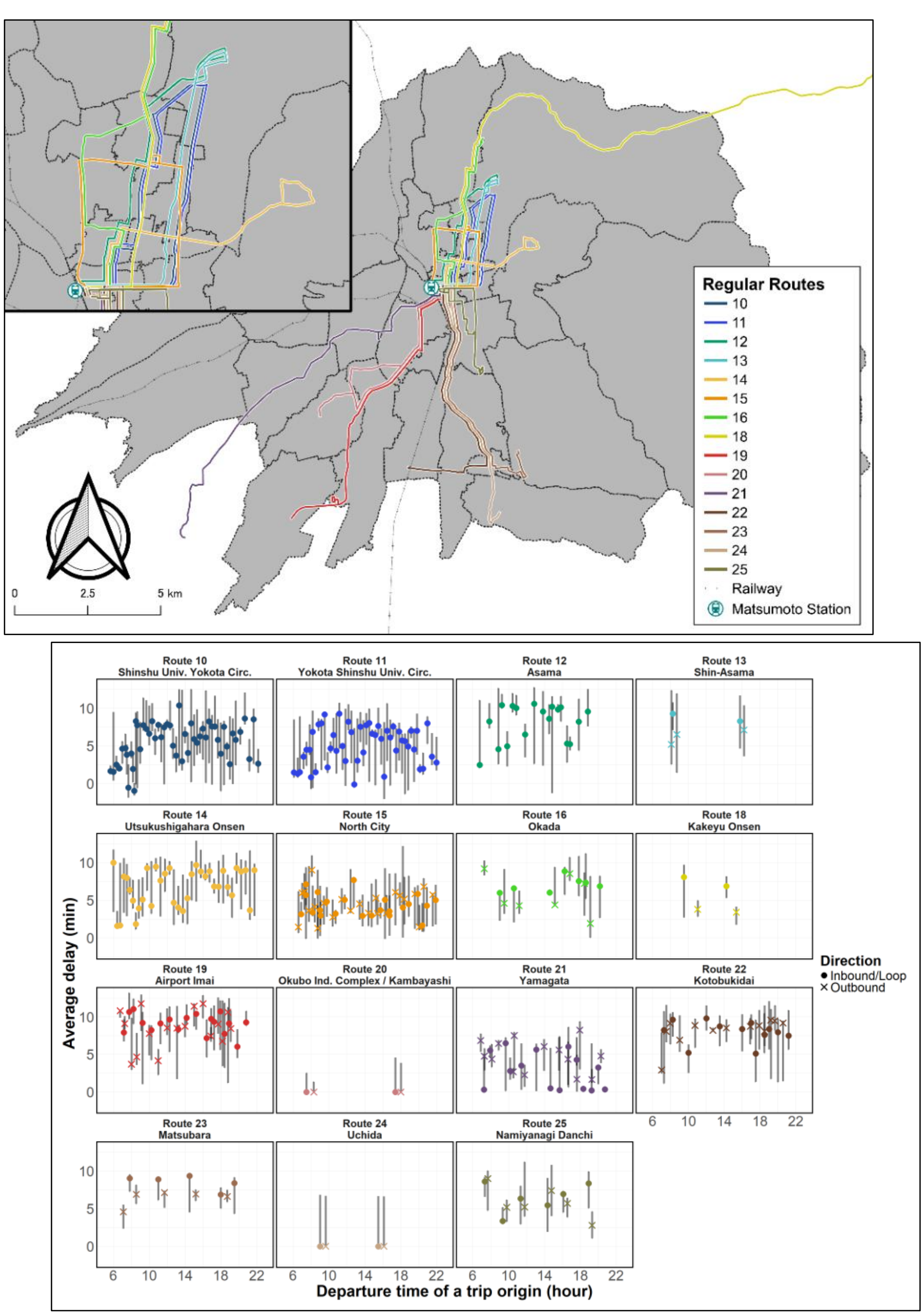

**Fig. 6** Color-coding of routes (top) and delay distribution of Regular Routes (except Routes 17 and 26: on 22-26 December 2025, bottom). Most routes except Routes 20 and 24 experience average delays of 5 minutes. The routes passing through the city center (Routes 10–15) exhibit delays up to 10 minutes.

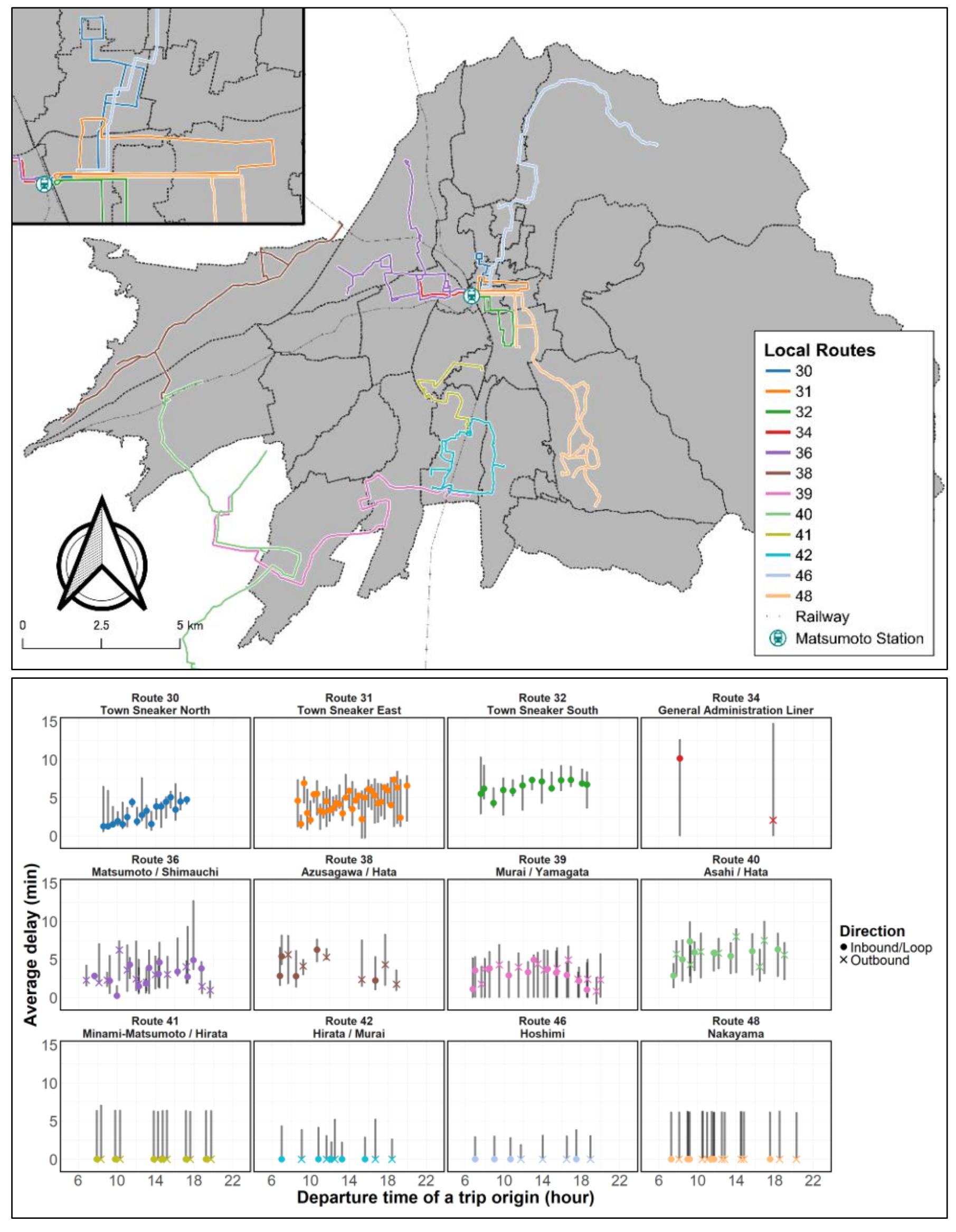

**Fig. 7** Color-coding of routes (top) and delay distribution of Local Routes (except Routes 33, 37, 43, 44, 45, and 47: on 22-26 December 2025, bottom). With the exception of the Town Sneaker routes (Routes 30-32) and Route 34, which operate through the city center, the average delay per service generally remains within the range of 0 to 5 minutes.

In addition, the railway and road network data required for executing was based on OpenStreetMap (OSM, https://www.openstreetmap.org/), which was retrieved from Geofabrik (https://www.geofabrik.de/), and subsequently processed to extract the network within the designated city planning area.

## 3.4 Data on travel time

### 3.4.1 Travel-time calculation

As previously stated, this study employs the r5r package in R. The package reads GTFS and OSM data to construct a transportation network and computes shortest routes and travel times for specified departure times. Departure time windows are divided into 30-minute intervals, within which travel times are calculated at one-minute intervals for each OD pair. For each window, two representative values—the median and the 25th percentile—are extracted. The 25th percentile reflects the assumption that students in regional cities with infrequent services plan their commutes broadly according to the published timetable (Stewart and Byrd 2023), while the median value is used only as a reference for timetable-based accessibility. For return trips, OD pairs are reversed, and departure times are adjusted, and travel times are calculated using the same procedure.

Furthermore, when estimating travel times under actual conditions, we placed particular emphasis on avoiding the overestimation of accessibility. In this study, we adopted an approach in which the departure time corresponding to the percentile derived from schedule-based travel times is used as the reference point, and travel times are then recalculated using r5r to reflect actual operating conditions. The detailed procedure is as follows.

First, a travel-time distribution under actual operating conditions is constructed using GTFS data that incorporate observed delays. From this distribution, the trip departing at the same time as the representative scheduled travel time (defined by the corresponding percentile under scheduled conditions) is extracted, and its travel time is adopted as $t_{ikw}$ or $t_{ikw'}$.

It is unrealistic to assume that travelers have perfect prior knowledge of the delay distribution and can choose their departure times optimally based on such information. By conditioning actual travel times on the scheduled departure time, the proposed approach more realistically represents traveler behavior, in which departure decisions are made according to the timetable without full anticipation of future delays.

Trips that do not arrive by the start of class (8:40 a.m.) are excluded from the analysis, and OD pairs with no valid travel time within the time windows $w$ and $w'$ for commuting to and returning from school are excluded. As shown in Equation (1), the round-trip travel time between cell centroid $i$ and school $k$, $t_{ik}$, which forms the basis of the accessibility index, is calculated by adding the weighted travel times for the round trips:

$$t_{ik} \coloneqq \sum_{w} p_w \cdot t_{ikw} + \sum_{w'} p_{w'} \cdot t_{ikw'} \tag{1}$$

where,

$w$: time window in commuting to school
$w'$: time window in returning from school
$p_w$: departure share in time window $w$
$p_{w'}$: departure share in time window $w'$
$t_{ikw}$: travel time from cell $i$ to facility $k$ in time window $w$
$t_{ikw'}$: travel time from facility $k$ to cell $i$ in time window $w'$

The departure share $p_w$, shown in Table 1, is based on the results of the *Matsumoto City Person Trip Survey* ('PT survey' hereafter) conducted in 2019. The two most significant hours were extracted from the time distribution of trips. Data shows that most trips to school and kindergarten occur between 7:00 a.m. and 9:00 a.m. (Matsumoto City, Yamagata Village and Asahi Village 2023). We changed the measurement period to 6:00-8:30 a.m. because of the start times of high schools in the city. For returning home from school, we divided the 4:00-8:00 p.m. time slot into eight 30-minutes slots. Here, it is also necessary to obtain the departure share $p_{w'}$ for the return trip, similar to the commuting trip. However, respondents of PT survey include many people with different attributes, such as adult commuters and shoppers, which makes it difficult to extract trips done exclusively by students. Therefore, we assume that return trips are evenly distributed across the eight time windows (i.e., $p_{w'} = 0.125$ for any $w'$). These sets of travel times are calculated for each of the five weekdays, incorporating the five-day delay data obtained in Subsection 3.3.2.

**Table 1** Departure shares in morning time window $p_w$

| Time window $w$ | 6:00-6:30 | 6:30-7:00 | 7:00-7:30 | 7:30-8:00 | 8:00-8:30 |
|---|---|---|---|---|---|
| Share | 0.041 | 0.041 | 0.396 | 0.396 | 0.126 |

### 3.4.2 Routing conditions

Table 2 presents a comprehensive list of conditions for routing using r5r. In the *Matsumoto Regional Public Transport Plan*, areas located beyond a 500-meter radius from bus stops or a 1-kilometer radius from railway stations are classified as Public Transport Deserts (Matsumoto City, Yamagata Village and Asahi Village 2023). In accordance with this plan, this study sets the maximum walking access time to bus stops at 7.5 minutes and to railway stations at 15 minutes.

The *Matsumoto Regional Public Transport Plan* also notes that many students across the city choose bicycles as one of their commuting modes (Matsumoto City, Yamagata Village and Asahi Village 2023). For the purpose of analyzing commuting accessibility via public transport, this study does not evaluate trips consisting entirely of bicycle travel. However, since most railway stations are equipped with bicycle parking facilities, cycle-and-ride trips from residential areas to stations are considered. This allows some grid cells within Public Transport Deserts to become partially accessible. The access time for these bicycle trips is set at 15 minutes, corresponding to the walking access limit. In contrast, cycle-and-ride trips to bus stops are excluded because bicycle parking is not available at most bus stops.

Additionally, when a trip involves bicycle use for the journey to school, a constraint is imposed such that the same station must be used for the return trip. These constraints allow the derivation of travel-time distributions based on feasible trips.

**Table 2** Routing conditions for calculating travel-time matrices and output indices

| Details | Values | Remarks |
|---|---|---|
| Residence points | 1,244 cells | Centroids of 250m grid in the city planning area |
| Facilities | 9 schools | General education high schools in the city |
| Population | 29,574 people | U15 population in the city planning area |
| Routing date | Dec. 22-26 2025 | Eliminating the effects of the holiday congestions |
| Walk/cycling speed | 4km/h/12km/h | 4km/h for walking, 12km/h for cycling |
| Number of transfers | Once (Max.) | – |
| Duration of access/egress walk | 7.5min./15min. (Max.) | Definition of PT Deserts:<br>7.5min. for bus stops, 15min. for stations<br>(15min.station access/egress by bicycle also accepted) |
| Duration of walk-only trips | 15min. (Max.) | Equivalent to access/egress duration<br>(except bicycle-only trips) |

## 3.5 Output indices

For calculating the accessibility indices, we define accessibility as "the availability of opportunities for citizens to participate in activities and return home." In this case, it can be rephrased as the ability of children in junior high school or younger to commute to high school and return home; we can evaluate the freedom of citizens to choose their educational path by measuring the potential opportunities for elementary and junior high school students to attend multiple high schools, not considering accessibility of high school students, who have already chosen which school to enroll. The availability of transportation in residential areas, or *System Accessibility* (Malekzadeh and Chung 2020), does not guarantee the ability to reach destinations or activities. In other words, while the commonly used transit stops coverage rate (Matsumoto City, Yamagata Village and Asahi Village 2023), calculated based on the coverage of stop buffers, is easy to calculate, it is an inappropriate accessibility indicator for regional cities with low service frequencies.

We introduce the concept of *Expected Cumulative Opportunities* (ECO, $ECO_i$) as a basic accessibility index for each grid cell $i$, as defined in Equations (2) and (3). In general, the accessibility of facility $k$ from cell $i$, denoted $c_{ik}$, is formulated as a function in which the attractiveness of the facility decreases with increasing travel time $t_{ik}$. The function $f(t_{ik})$ in Equation (3) can be interpreted as the proportion of people who can tolerate, or the probability that a person tolerates, the travel time $t_{ik}$, taking values between 0 and 1.

For convenience, in this study, $f(t_{ik})$ is defined as 1 when $t_{ik}$ is shorter than the travel-time threshold $T$, 0 when $t_{ik}$ exceeds $2T$, and decreases linearly with $t_{ik}$ between these limits. Then, $ECO_i$ is calculated as the sum of $c_{ik}$ across all schools, representing the expected total attractiveness of facilities accessible within the acceptable travel time. This allows the intuitive interpretation that ECO can be regarded as the equivalent number of schools accessible within the travel-time threshold $T$. Facilities located farther than $2T$ are excluded from the set of accessible opportunities.

$$ECO_i \coloneqq \sum_k c_{ik} \tag{2}$$

$$c_{ik} \coloneqq f(t_{ik}) = \begin{cases} 1, & t_{ik} \leq T \\ 2 - \dfrac{t_{ik}}{T}, & T < t_{ik} \leq 2T \\ 0, & 2T < t_{ik} \end{cases} \tag{3}$$

where,

$T$: travel-time threshold for round-trip
$c_{ik}$: accessibility of facility $k$ for cell $i$

Regarding the setting of the travel-time threshold $T$, we employed three values for a sensitivity analysis. According to the *2021 Survey on Time Use and Leisure Activities*, high school students in Nagano Prefecture spend an average of 94 minutes on round-trip commuting (Statistics Bureau of Japan 2022). Based on this, the thresholds $T$ were set at 60 min (stricter threshold), 90 min (moderate threshold), and 120 min (more lenient threshold).

We also conducted a district-level analysis to identify populations benefiting from accessibility. The population of each district was calculated by aggregating the populations of grid cells that intersect with or are contained within the district boundaries. To enable a comparative assessment of ECO across districts, we computed population-weighted ECO indicators using the median and the 75th percentile, representing the ECO values below which 50% and 75% of the district's population fall, respectively. This approach captures accessibility experienced by the average student and the majority of students in each district, rather than relying on unweighted median ECO values.

The discrepancy of opportunities between the timetable-based and actual operational scenarios is quantified as $OGL_i$ (Opportunity Gain/Loss), defined as the difference in ECO between normal and delayed conditions. A negative $OGL_i$ indicates that opportunity loss occurs due to delays. Additionally, $OGL_i$ (Total Opportunity Gain/Loss) is calculated as the product of $OGL_i$ and the population of each grid cell, representing the demographic impact of opportunity gains and losses caused by bus delays. Both indices are calculated following Equations (4) and (5).

$$OGL_i \coloneqq ECO_{i,actual} - ECO_{i,scheduled} \tag{4}$$

$$TOGL_i \coloneqq N_i \cdot OGL_i \tag{5}$$

where,

$N_i$: (U15) population of cell $i$

Note that all the indices are calculated based on travel times computed separately for each weekday. For figures and tables, median values over the five weekdays are presented as representative values.

# 4 Accessibility results under scheduled networks

## 4.1 Analysis by 250m grid

Before examining the distribution of accessibility losses due to delays, this section presents accessibility under scheduled public transport networks in order to clarify spatial disparities in students' school choice opportunities. Table 3 shows the population distribution of ECO based on an optimistic percentile (25th percentile) of travel time within the time window, assuming that students align their commuting schedules closely with published timetables. In addition, as a supplementary analysis, the frequency distribution of ECO based on the median travel time, representing average accessibility conditions, is also presented.

Under a 60-minute round-trip travel-time threshold, when optimistic commuting conditions are assumed, 17.8% of students live in highly accessible areas where five or more high schools can be reached. Meanwhile, 31.5% of students fall into a moderate accessibility category, with access to between two and fewer than five schools, indicating that nearly half of all students are already able to commute to multiple high schools. By contrast, 32.5% of the population has an ECO value of less than 1, suggesting that approximately one-third of the students in the city must rely on non-public transport options—such as cycling or being driven by car—for either one-way or round-trip commuting.

When the round-trip travel-time threshold is extended to 90 minutes, the population benefiting from accessibility increases sharply. The proportion of the population able to commute to at least one high school rises from 67% to 78%, representing an increase of 11 percentage points. In particular, the share of students able to commute to seven or more high schools under optimistic conditions increases substantially, from 4% to 38.9%. This threshold corresponds to the average commuting time of high school students in the city and thus represents a realistically attainable level of accessibility.

In contrast, relaxing the round-trip travel-time threshold to 120 minutes yields only limited improvements in accessibility. While the use of median travel times results in an increase in the population able to commute to seven or more high schools, under optimistic conditions the share of the population able to reach at least one high school increases only marginally, from 78% to 80%. Moreover, approximately 17% of the population still needs to rely on cycling or escort by private car for at least one leg of the trip, indicating that further relaxation of the time constraint plays a limited role in expanding commuting opportunities.

**Table 3** ECO population distribution by travel-time threshold. PT Deserts follows the definition by Matsumoto City.

| | $T$ = 60 (min.) | | $T$ = 90 (min.) | | $T$ = 120 (min.) | |
|---|---|---|---|---|---|---|
| | 25th perc. | Median | 25th perc. | Median | 25th perc. | Median |
| $7 \leq ECO \leq 9$ | 1,383 (4.7%) | 656 (2.2%) | 11,511 (38.9%) | 7,192 (24.3%) | 18,122 (61.3%) | 16,300 (55.1%) |
| $5 \leq ECO < 7$ | 3,862 (13.1%) | 2,557 (8.6%) | 5,322 (18.0%) | 6,883 (22.6%) | 1,298 (4.4%) | 2,945 (10.0%) |
| $2 \leq ECO < 5$ | 9,311 (31.5%) | 8,174 (27.6%) | 5,018 (17.0%) | 7,266 (24.6%) | 2,538 (8.6%) | 2,583 (8.7%) |
| $1 \leq ECO < 2$ | 5,420 (18.3%) | 3,799 (12.8%) | 1,200 (4.1%) | 1,001 (33.9%) | 1,973 (6.7%) | 1,844 (6.2%) |
| $0.5 \leq ECO < 1$ | 1,592 (5.4%)) | 4,241 (14.3%) | 1,097 (3.7%) | 1,717 (5.8%) | 430 (1.5%) | 689 (2.3%) |
| $0 \leq ECO < 0.5$ | 8,006 (27.1%) | 10,147 (34.3%) | 5,426 (18.3%) | 5,515 (18.6%) | 5,213 (17.6%) | 5,213 (17.6%) |
| PT Deserts | | | 3,368 (13.0%) | | | |

Based on these results, this study identifies a 90-minute round-trip travel-time threshold as a representative for evaluating school commuting opportunities. Figure 8 illustrates the median ECO values calculated over five days, assuming a scheduled (non-delayed) network and optimistic commuting behavior under the 90-minute threshold. Spatial distributions of accessibility for other thresholds are provided in Appendices 3 and 4.

According to the city's definition, Public Transport Deserts are home to 3,368 U15 students (13.0% of all the students). In these areas, it is generally difficult to commute to schools only by public transport and on foot; however, since this definition does not account for the possibility of cycle-and-ride trips, as Figure 8, some areas inside Public Transport Deserts have non-zero ECO values.

Accessibility is notably high along railway corridors, particularly the JR Shinonoi Line Corridor and the Kamikochi Line Corridor, where grid cells allowing commuting to and from multiple high schools are widely distributed. In addition, high ECO values (seven or more schools) are observed in the Central Area Buffer surrounding Matsumoto Station. In the Iriyamabe District, adjacent to the eastern side of the Central Area Buffer, hatched grid cells indicate weekday-dependent variations in accessibility caused by alternating-day service, even under scheduled operations. Beyond the major railway corridors, relatively high accessibility is also observed in cells served by Regular Routes bus services. In contrast, Public Transport Deserts are concentrated in the peripheral areas of the city. In most of these cells, ECO values remain below 0.5, and cycle-and-ride options are largely ineffective, indicating a near-total reliance on cycling or parental escort (by private car) for school commuting.

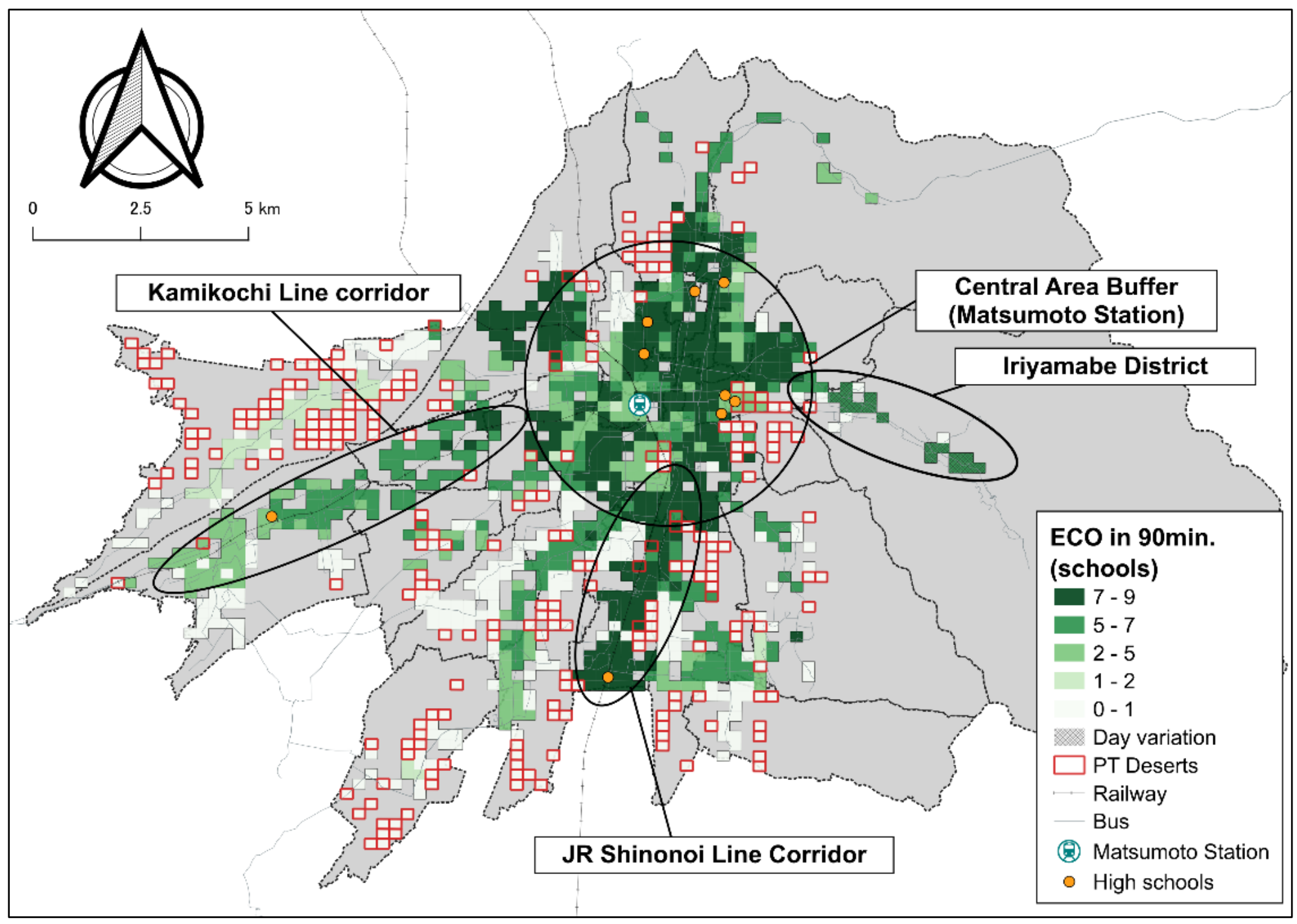

**Fig. 8** Spatial distribution of ECO in 25th percentile value of travel time, in round trip thresholds of $T$ = 90 (min.)

## 4.2 Analysis by district

This subsection examines the level of accessibility enjoyed by the majority of the population in each district. Specifically, as outlined in Subsection 3.5, ECO values at the 50th and 75th population percentiles within each district are evaluated, taking grid-level population distribution into account. Here, travel-times are based on the 25th percentile, assuming optimistic commuting conditions in which students broadly align their schedules with published timetables. Figure 10 presents the plots of the U15 population and district-level ECO values under three travel-time thresholds (more detailed values are shown in Appendices 5 and 6). The upper panel shows trends based on the 50th population percentile of the district population, while the lower panel shows those based on the 75th population percentile. Circles, triangles, and squares represent the different travel-time thresholds defined in Subsection 3.5, and the clusters formed by connecting these symbols illustrate changes in accessibility within individual districts.

Focusing on the 50th percentile of the population, an expansion of the round-trip travel-time threshold from 60 to 120 minutes results in an average increase of approximately 4.654 in ECO across all districts. In other words, allowing a 45-minute increase in one-way commuting time—from 15 to 60 minutes—enables students to gain, on average, approximately four additional school options in terms of expected value.

In the five areas near Matsumoto Station—Areas of the Central, the Central North, the Central South, the Higashiyama North, and the Higashiyama Middle—most districts already have multiple commuting options within a 60-minute round-trip threshold, supported by well-developed public transport (Figure 10, upper panel). As a result, extending the threshold yields only limited gains, with an average ECO increase of 4.206 from 60 to 120 minutes, below the overall average.

In the four areas of the Hata, Kasai North, Kasai South, and South Areas, accessibility at the 60-minute threshold is limited to one or two schools but increases sharply as the constraint is relaxed. The average ECO increase reaches 5.856, exceeding that of other areas.

In contrast, some districts in the Higashiyama South and Azusagawa areas have ECO values below 1, depriving minimum necessary access to educational opportunities. Accessibility gains remain minimal even with relaxed constraints (average increase: 0.389). The Azusagawa Area, with over 1,000 students, is particularly affected. Grid-level analysis suggests intra-district disparities, with access concentrated near high schools and bus routes, highlighting the need for targeted public transport improvements.

Under the 75th percentile condition, even high-accessibility areas such as the Central North and Central South areas include districts that cannot access multiple high schools within a 60-minute round-trip. In the South Area, where improvements were seen at the 50th percentile, some districts still have ECO values of one or less even at 120 minutes. This suggests that accessibility remains constrained for populations between the 50th and 75th percentiles.

The preceding analysis examined accessibility outcomes at the district level while accounting for population distribution. However, under actual operational conditions, accessibility as assumed from scheduled timetables may be substantially degraded. In the following sections, we, therefore, analyze localized accessibility losses using a grid-based approach in order to investigate the impacts of real-world operational conditions on commuting accessibility.

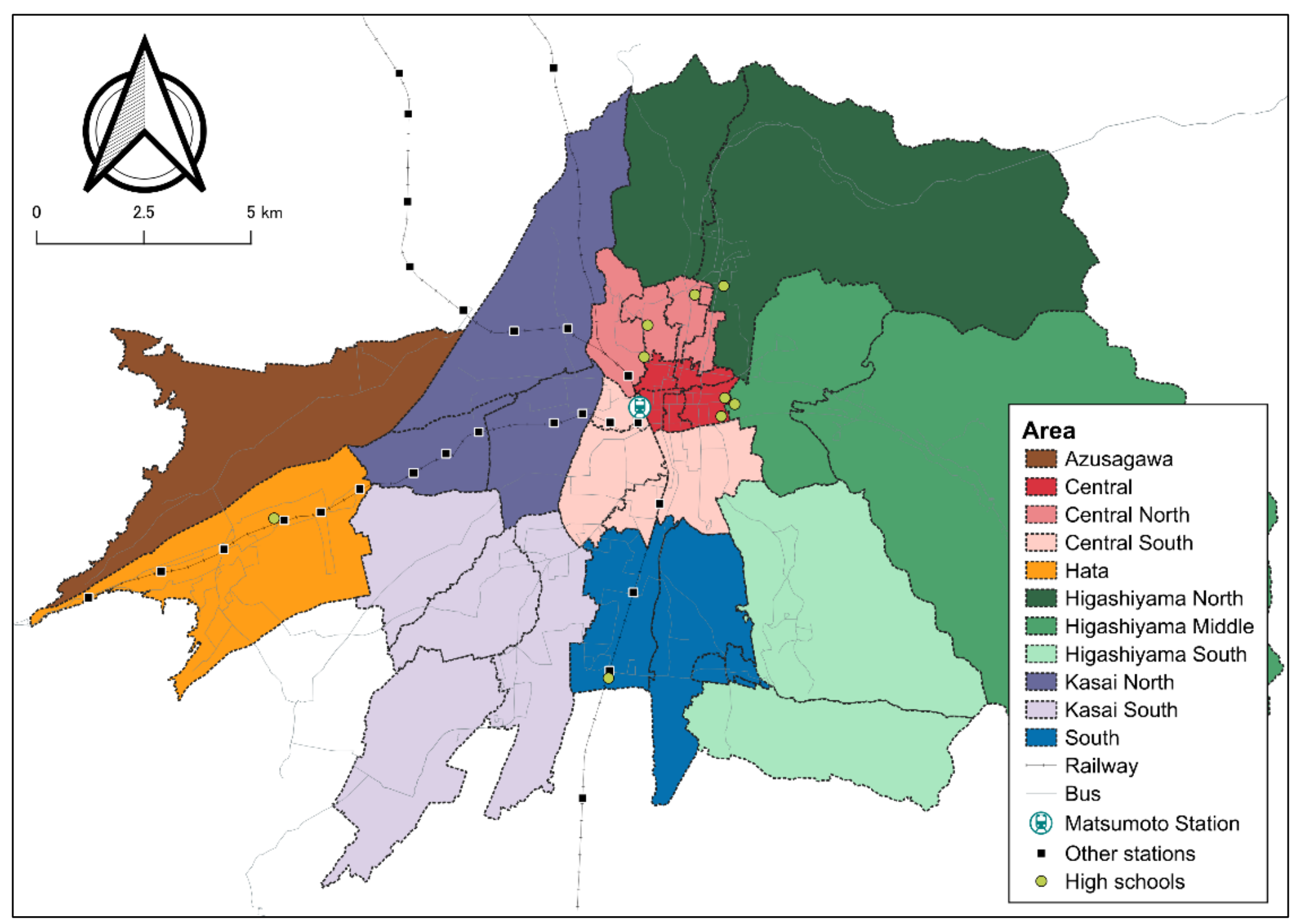

**Fig. 9** Color-coding and locations of areas for a district level analysis

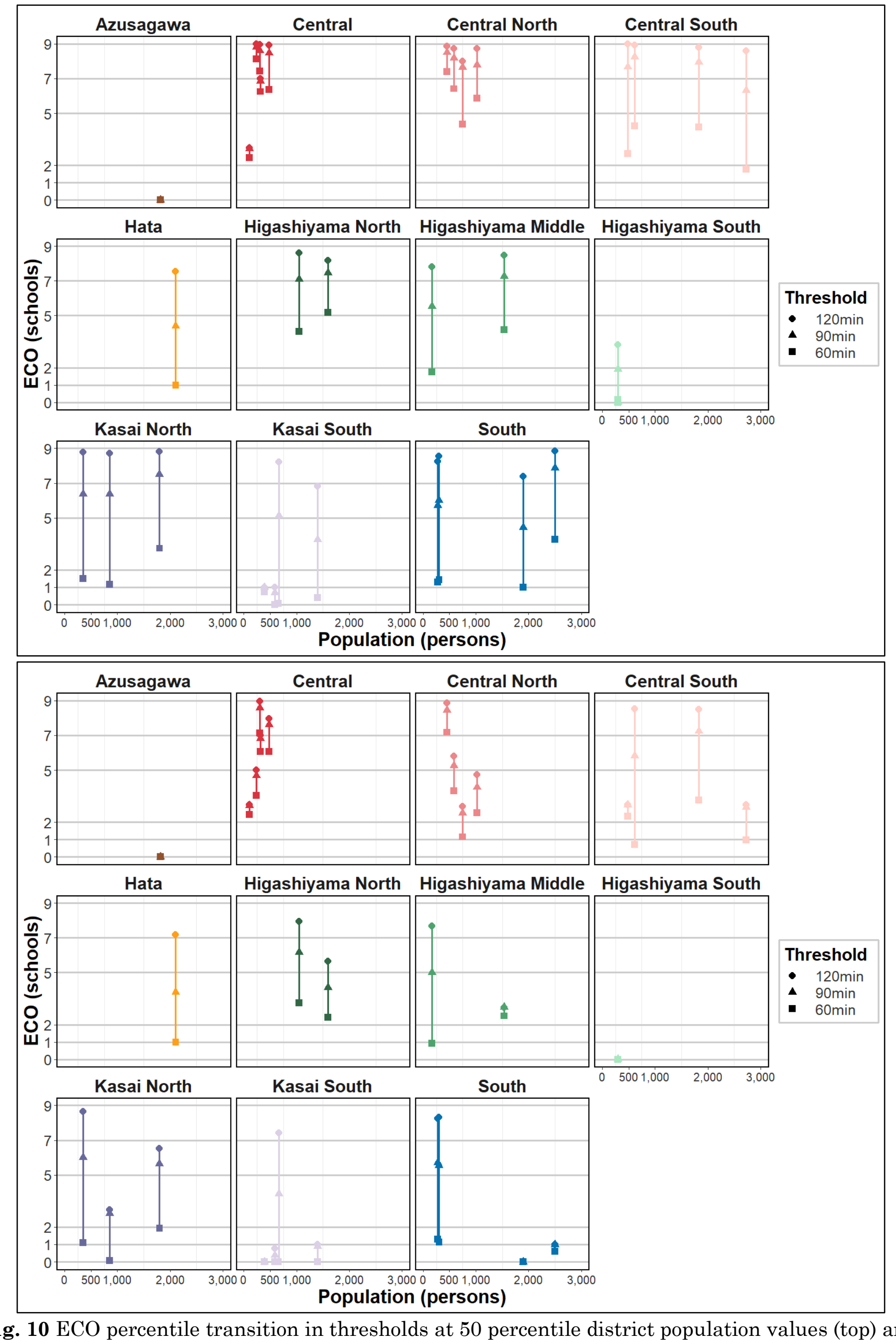

**Fig. 10** ECO percentile transition in thresholds at 50 percentile district population values (top) and 75 percentile population values (bottom)

# 5 Accessibility results under actual conditions

## 5.1 Spatial inequality in ECO

In this section, we show the results on the accessibility considering operational conditions, compared to timetable-based metrics discussed in the previous section.

First, for each travel-time threshold, we calculated the Gini coefficient of ECO across all grid cells in the study area, weighted by the population under 14 years of age, using the median ECO over five weekdays. Table 4 presents the Gini coefficients for ECO in both the scheduled timetable network and the network reflecting actual operational conditions. The values in parentheses indicate the 95% confidence intervals obtained from 1,000 bootstrap iterations. Coefficient values suggest significant disparities across districts in terms of ECO. Regarding how does accounting for delays affect such disparities, although the Gini coefficients exhibit a slight worsening under delay conditions across all travel-time thresholds, none of these differences are statistically significant.

**Table 4** Gini index of ECO by travel-time threshold

| | $T$ = 60 (min) | $T$ = 90 (min) | $T$ = 120 (min) |
|---|---|---|---|
| Scheduled | 0.515<br>(CI: 0.494-0.537) | 0.363<br>(CI: 0.339-0.387) | 0.320<br>(CI: 0.293-0.347) |
| Actual | 0.562<br>(CI: 0.538-0.582) | 0.404<br>(CI: 0.379-0.426) | 0.357<br>(CI: 0.329-0.382) |

## 5.2 Opportunity Gain/Loss (OGL) due to delay

The distribution of OGL is presented in Table 5 and Figure 11. Because the spatial patterns of accessibility gains and losses due to delays differ substantially depending on the travel-time threshold, distribution maps for all thresholds $T$ are shown in this section. Note that all the OGL and TOGL values referred to in these subsections are based on the 25th percentile travel time.

Under the 60-minute threshold, delays result in accessibility losses in 706 cells (57% of the study area), particularly along bus corridors in the JR Shinonoi Line and the Central Area Buffer. Limited service frequency implies that delays directly increase travel times, especially on routes such as 10, 11, 19, 22, and 32. In contrast, localized accessibility gains are observed in some areas, such as the Niimura District, due to reduced waiting times.

At the 90-minute threshold, accessibility losses along railway corridors become less pronounced, whereas losses of one or more OGL loss in parts of the South and Kasai South areas. This suggests reduced sensitivity to delays in rail-served areas, while bus-dependent areas experience diminished accessibility.

Overall, the number of unaffected cells slightly decreases, accompanied by a modest increase in cells with opportunity losses.
At the 120-minute threshold, OGL values along railway corridors converge toward zero, as most high schools become reachable within 60 minutes one-way. In contrast, areas with limited baseline accessibility continue to exhibit persistent opportunity losses.

While the discussion thus far has focused on opportunity losses, a notable finding is that accessibility improves under delays in the Higashiyama North Area across all travel-time thresholds. While the Niimura District at $T = 60$ and the Satoyamabe District at $T = 120$ are enabled to reduce waiting time for boarding buses and transfer in the same routes as the scheduled, those improvements in the Higashiyama North Area are mainly attributable to *Lucky Catch* events. Lucky Catch refers to a situation in which a delay enables a student to board a connecting bus that would have been missed under scheduled operations. Figure 12 illustrates the differences between scheduled and delayed routes departing from a specific location in the two districts. The figure clearly shows that, for some high schools that would normally require a detour, the availability of an additional transfer results in arrival times that are approximately 10 minutes earlier. Nevertheless, even in these districts, the increase in opportunities does not exceed one, meaning that the equivalent increase in commuting opportunities—based on round-trip travel-time thresholds of 60, 90, and 120 minutes—remains below one school. Therefore, the extent to which such delay-induced opportunity gains are meaningful for students warrants

**Table 5** OGL distribution of the number of cells by travel-time threshold

| | $T = 60$ (min.) | | $T = 90$ (min.) | | $T = 120$ (min.) | |
|---|---|---|---|---|---|---|
| | N. of cells | (%) | N. of cells | (%) | N. of cells | (%) |
| $0 < OGL < 1$ | 28 | (2.3%) | 30 | (2.4%) | 39 | (3.1%) |
| $OGL = 0$ | 88 | (7.1%) | 84 | (6.8%) | 221 | (17.8%) |
| $-1 \leq OGL < 0$ | 453 | (36.4%) | 489 | (39.3%) | 425 | (34.2%) |
| $-3 \leq OGL < -1$ | 240 | (19.3%) | 252 | (20.3%) | 169 | (13.4%) |
| $OGL \leq -3$ | 13 | (1.0%) | 33 | (2.7%) | 34 | (2.7%) |
| Unreachable | 356 | (28.6%) | 356 | (28.6%) | 356 | (28.6%) |
| Max. OGL | 0.559 | | 0.283 | | 0.144 | |
| Min. OGL | -4.809 | | -6.675 | | -7.739 | |

According to the OGL analysis results, as the travel-time threshold increases, the absolute magnitude of commuting opportunity losses or gains generally tends to increase. In contrast, along highly punctual railway corridors, the absolute values of OGL tend to converge toward zero. In other words, both accessibility losses and gains induced by delays become more pronounced with relaxed time constraints, with accessibility losses being the most common outcome. It should be noted that in areas where no high schools are reachable by public transport, OGL is by definition zero.

The improvements observed in terms of accessibility under actual operating conditions can be explained primarily by Lucky Catch events and by simple reductions in waiting time. However, in such routes and areas, an undesirable situation arises which adhering strictly to the published timetable results in poorer outcomes for travelers. This finding suggests the need to revise timetables based on actual operational performance.

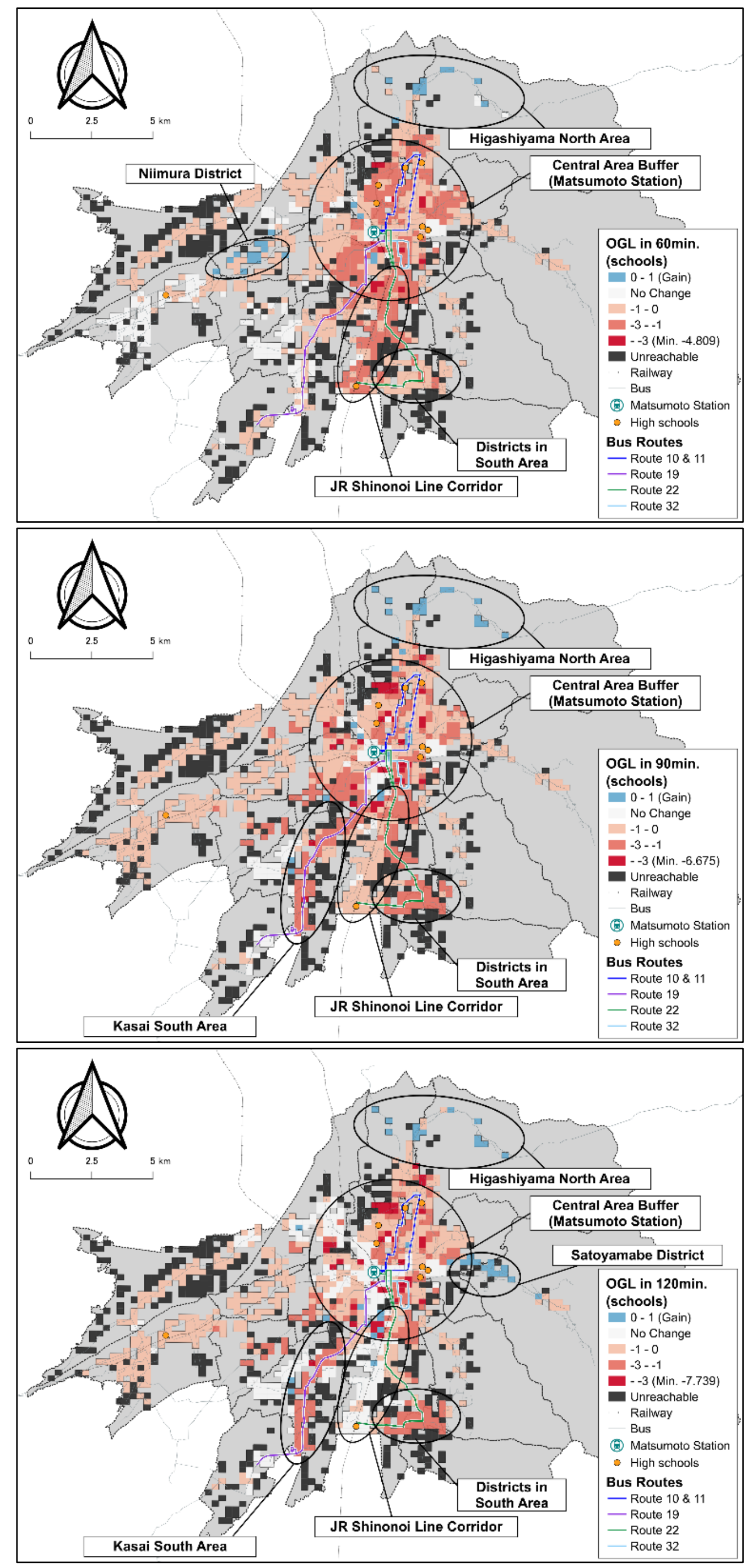

**Fig. 11** OGL in round trip thresholds of $T$ = 60min. (top), 90min. (center), and 120min. (bottom)

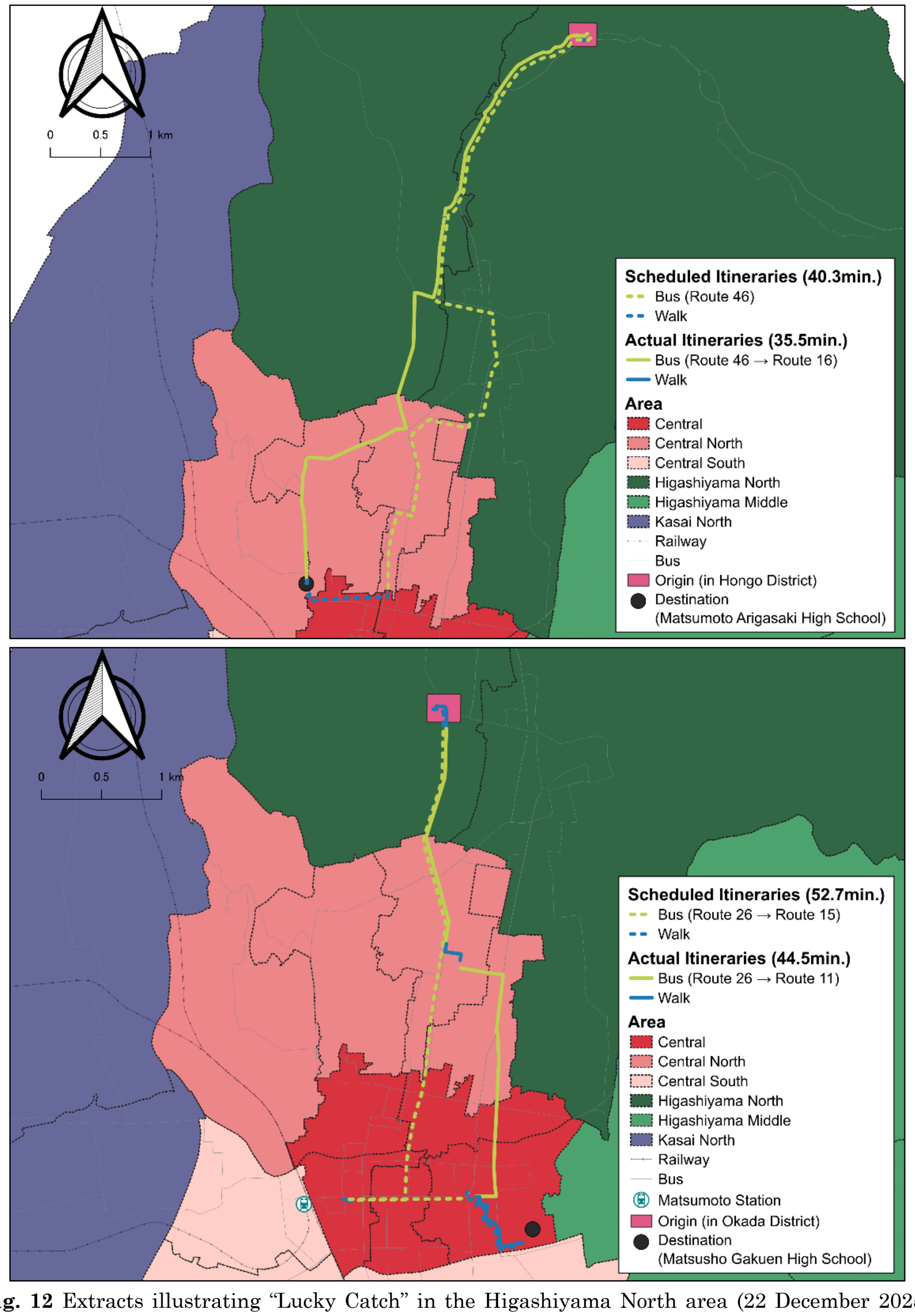

**Fig. 12** Extracts illustrating "Lucky Catch" in the Higashiyama North area (22 December 2025). Dotted lines indicate scheduled services, while solid lines represent actual operations. In the upper panel, students were able to make an additional transfer to another route under actual operating conditions. In the lower panel, an alternative route became available midway through the scheduled routes.

## 5.3 Total Opportunity Gain/Loss (TOGL) due to delay

The TOGL index (see Figure 13) was introduced to evaluate changes in accessibility while accounting for population size. Unlike ECO and OGL, its values cannot be interpreted as the expected number of accessible school opportunities from residential grid cells; however, it enables the identification of areas where accessibility losses affect a larger population. As with OGL, the absolute value of TOGL tends to increase as the travel-time threshold is relaxed across all thresholds.

In comparison with the OGL distribution shown in Figure 11, accessibility losses observed in districts within the South Area under the 90-minute and 120-minute travel-time thresholds exhibit particularly large population-weighted impacts due to high grid-level populations. In particular, severely affected areas with TOGL values below −100 are observed around the Kotobukidai and Matsubara Districts. These areas are served by Route 23, one of the regular bus routes; however, services are limited to a small number of trips during the morning and evening peak periods, and persistent delays of approximately five minutes are common. As a result, students commuting by public transport in these districts are almost certain to experience delays, leading to substantial inconvenience. When considering service improvements such as timetable restructuring, it is, therefore, essential to account for the demographic impacts of accessibility losses caused by delays.

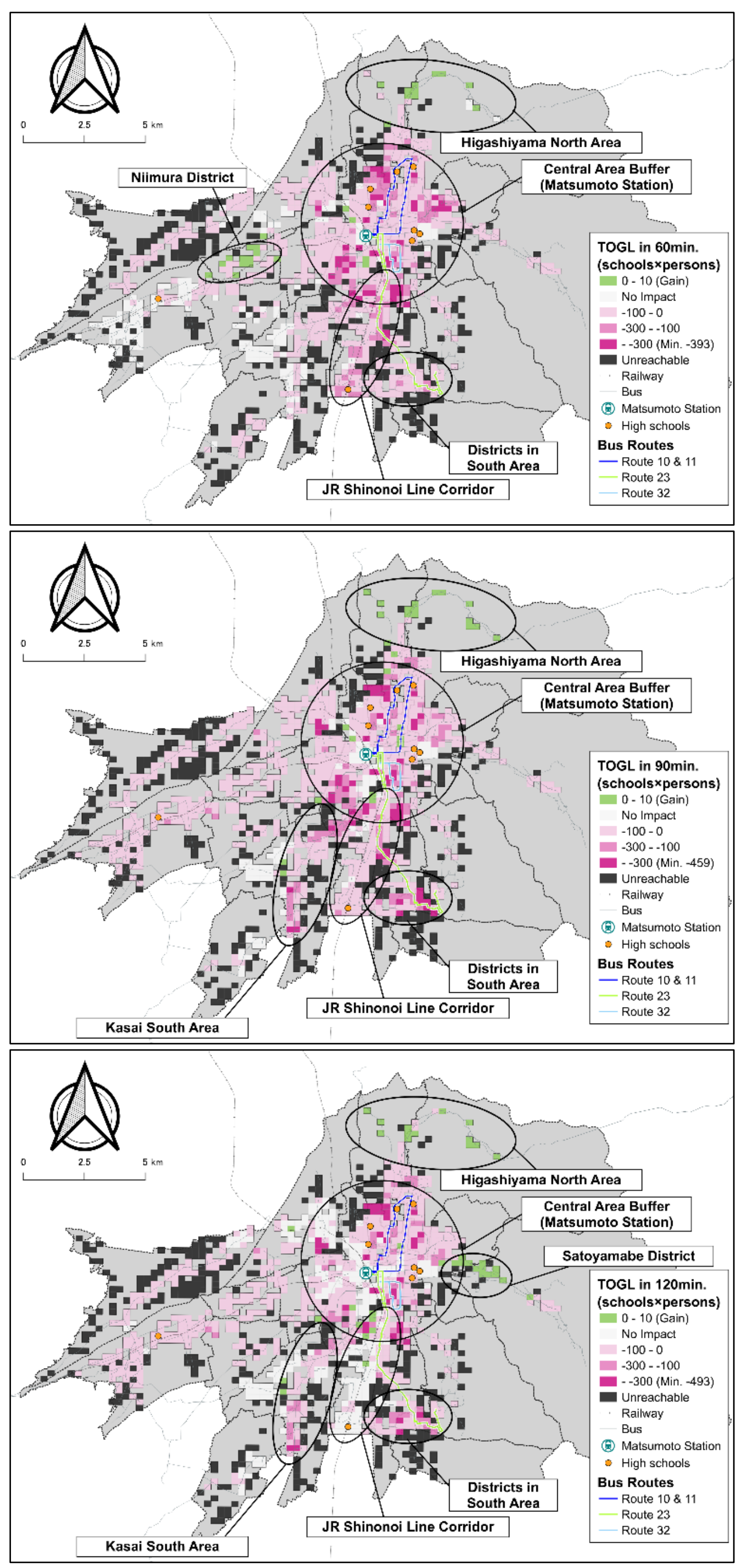

**Fig. 13** TOGL in round trip thresholds of $T$ = 60min (top), 90min (middle), and 120min (bottom)

# 6 Conclusion

## 6.1 Discussions of results and policy implications

In this study, we evaluated public transport accessibility to high schools in Matsumoto City by calculating travel times using two types of GTFS data: one reflecting actual operating conditions over five consecutive days, including delays, and the other based on a predefined timetable. Travel times were calculated for both outbound and return trips, taking into account access modes and cycle-and-ride options, in order to represent realistic public transport use. In addition, a sensitivity analysis was conducted for the proposed accessibility indices. All indices employed in this study are cumulative measures, and their output values can be interpreted as the expected number of accessible commuting opportunities. This characteristic overcomes the conventional problem of accessibility being represented as a dimensionless and unintuitive quantity, thereby enabling more intuitive interpretation (Kapatsila et al. 2023).

Regarding the validity of incorporating round-trip accessibility into the evaluation, return trips in Matsumoto City generally have lower service frequency than outbound trips. However, because travel time is influenced by timetables and actual operating conditions, considering the return trip may either increase or decrease waiting times, making the evaluation more or less stringent; thus, no uniform conclusion can be drawn. Nevertheless, compared with evaluations based solely on one-way trips, incorporating round trips is considered to provide a more impartial assessment.

The grid-based analysis using Expected Cumulative Opportunities (ECO) revealed that, at a 60-minute travel-time threshold, 20% of residential grid cells had no feasible access to any high school for commuting to and from school. Even when a longer round-trip threshold of 120 minutes—exceeding typical average commuting times—was applied, approximately one-third of the residential cells and about one-tenth of the students were still unable to complete a round trip to any high school. District-level ECO analysis showed that relaxing the round-trip threshold to 120 minutes allowed more than 50% of the students in almost all districts to access at least one high school. However, more than 75% of the students in 7 out of 32 districts had no access to even a single high school. In particular, Azusagawa District, which has a large area and a high U15 population, contains extensive public transport service gaps; while access to at least one high school is possible along bus corridors, accessibility is highly uneven within the district.

These findings underscore the limitations of assessing residents' convenience using simple indicators such as distance to bus stops alone. Furthermore, analyses using Opportunity Gain/Loss (OGL) and Total Opportunity Gain/Loss (TOGL) showed that as students' tolerated travel-time thresholds increase, the magnitude of accessibility loss also grows, thereby making the potential travel time increases caused by delays more visible. Conversely, in northern areas such as the Higashiyama North Area, accessibility gains were also observed, where delays enabled unexpected transfers that ultimately reduced total travel time. These effects are particularly pronounced in areas that rely on bus-based public transport along corridors located slightly away from the city center, highlighting the need for improvements to public transport timetables in such regions. The results provide a clear example of how accessibility based on actual operating conditions can substantially diverge from timetable-based evaluations, representing a key contribution that distinguishes this study from existing research in Japan.

Regarding the so-called *Lucky Catch* events, as shown in Figure 12, such events can in fact "enhance" accessibility in specific locations. However, for passengers to take advantage of this situation proactively—rather than remaining a matter of chance—the degree of delay must be relatively stable from day to day, enabling users to perceive these patterns and adjust their behavior accordingly. In other words, accessibility measures that fully incorporate actual operating conditions do not necessarily correspond to the accessibility that passengers experience in practice, because users often cannot accurately perceive delays or actual departure and arrival times.

Compared with previous accessibility analyses that incorporate delays, this study differs in several important aspects. While Liu et al. (2023) focused primarily on catchment areas, this study explicitly specifies activity opportunities and incorporates round-trip travel times, enabling a more realistic assessment of accessibility. Nichols et al. (2024), by contrast, analyzed accessibility to employment, where the number of opportunities is much larger than in the case of commuting to school, resulting in relatively higher accessibility values. Although the scope of opportunities examined in this study is more limited, the use of a distance-decay–based cumulative opportunity measure avoids the abrupt exclusion of opportunities caused by threshold-based approaches. This characteristic is particularly effective for evaluating school commuting accessibility, where the number of available destinations is inherently limited.

This study offers two main policy implications for both education and public transport planning. First, from a short-term perspective, it is essential to acknowledge the realities of school commuting. While reliance on cycling or parental escort by private car in areas with public transport service gaps may be unavoidable to some extent, measures such as adjusting school start times might be necessary to ensure that students can commute without excessive dependence on individual effort. Given the increase in private vehicle use during morning hours driven by commuting demand, bus prioritization measures—such as the introduction of exclusive bus lanes—or timetable adjustments aligned with actual operating conditions would be effective. In addition, the establishment of dedicated school commuting bus routes may serve as a promising strategy to improve accessibility in specific areas. Second, from a long-term perspective, when school consolidation becomes unavoidable in the future, accessibility by public transport should be incorporated as a key criterion in decision making. Although this study does not consider cycling-only commuting as an option, further discussion is needed regarding the extent to which public transport can realistically support school commuting, as well as whether a proactive shift from current commuting practices toward greater reliance on public transport should be encouraged.

## 6.2 Limitation and future work

While this study provides useful insights, a number of limitations point to avenues for further investigation. A potential research avenue is to establish a method for constructing GTFS data that incorporates expected delays, based on representative values of observed delay times obtained through the systematic and continuous collection of reliable operational data over a period of at least one month. In addition, the decay function used in accessibility indices could be refined by accounting for students' perceptions of travel time, for example by replacing linear functions with logistic functions. Ideally, such functions should be calibrated using empirical data. It is also important to examine whether incorporating information on actual arrival and departure times into timetable redesign could effectively enhance accessibility and provide an institutional response to the Lucky Catch situations.

In this study, all high schools were treated as equivalent in order to evaluate access to general academic high schools within the city. In practice, however, substantial differences exist among high schools in terms of educational characteristics, including curricular focus, extracurricular activities, and pathways to higher education or employment. Future research should, therefore, cluster high schools based on these characteristics and reassess accessibility from the perspective of the types of schools that students are able to attend.

In addition, it is important to examine how differences in accessibility are perceived by students as part of the attractiveness of potential school choices. For example, in the catchment area of Hachimori Junior High School, 45.6% of students reported that commuting accessibility influenced their high school choice (Matsumoto City, Yamagata Village and Asahi Village 2023). This school is located in the Imai District, which exhibits lower accessibility indices as shown in Figure 9 (Subsection 4.2), more specifically near the boundary between Matsumoto City and two neighboring villages. Future studies should explicitly consider how improvements in accessibility affect students' educational choices.

Finally, evaluating public transport performance should not be limited to accessibility to high schools, but should also include assessments for elderly populations who are unable to rely on private cars. In particular, examining accessibility during daytime hours—when students are generally not traveling—can reveal unexpected strengths or shortcomings of public transport systems that may not be apparent from analyses focused solely on school-related travel.

# Appendix 1 Frequencies of Regular Routes

| Route No. | Route name and section | Number of services | |
|---|---|---|---|
| | | Inbound/E. | Outbound/W. |
| **10** | **Shinshu Univ. Yokota Circulation Line** | **51** | **–** |
| | *- Loop* | *51* | *–* |
| **11** | **Yokota Shinshu Univ. Circulation Line** | **48** | **–** |
| | *- Loop* | *48* | *–* |
| **12** | **Asama Line** | **18** | **–** |
| | *- Loop* | *18* | *–* |
| **13** | **Shin-Asama Line** | **3** | **2** |
| | *- Matsumoto Sta. – Asama Onsen* | *3* | *2* |
| **14** | **Utsukushigahara Onsen Line** | **35** | **–** |
| | *- Loop* | *35* | *–* |
| **15** | **North City Line** | **23** | **23** |
| | *- Loop*<br>*- Matsumoto Sta. – Shinshu Univ. Hospital* | *6*<br>*17* | *6*<br>*17* |
| **16** | **Okada Line** | **7** | **7** |
| | *- Matsumoto Bus Term. – Yamashiroguchi* | *7* | *7* |
| **17** | **Alps Park Line** | **2** | **2** |
| | *- Matsumoto Bus Term. – Alps Park* | *2* | *2* |
| **18** | **Kakeyu Onsen Line** | **2** | **2** |
| | *- Matsumoto Bus Term. – Kakeyu Onsen* | *2* | *2* |
| **19** | **Airport Imai Line** | **16** | **16** |
| | *- Matsumoto Bus Term. – Shimo-Imai (Airport)* | *6 (10)* | *6 (10)* |
| **20** | **Okubo Factory Complex / Kambayashi Line** | **3** | **3** |
| | *- Matsumoto Bus Term. – Kambayashi Branch Office S.*<br>*- Matsumoto Bus Term. – Kambayashi Expwy. Stop* | *2*<br>*1* | *2*<br>*1* |
| **21** | **Yamagata Line** | **14** | **14** |
| | *- Matsumoto Bus Term. – Depot (Sasabe Danchi)* | *10 (4)* | *10 (4)* |
| **22** | **Kotobukidai Line** | **13** | **12** |
| | *- Matsumoto Bus Term. – Murai Sta. (Koike Crossing)* | *11 (2)* | *11 (1)* |
| **23** | **Matsubara Line** | **5** | **5** |
| | *- Matsumoto Bus Term. – Tanamine* | *5* | *5* |
| **24** | **Uchida Line** | **2** | **2** |
| | *- Matsumoto Bus Term. – Kuramura* | *2* | *2* |
| **25** | **Namiyanagi Danchi Line** | **6** | **6** |
| | *- Matsumoto Bus Term. – Namiyanagi Danchi* | *6* | *6* |
| **26** | **Shiga Line** | **6** | **6** |
| | *- Matsumoto Bus Term. Shiga Branch Office* | *6* | *6* |

# Appendix 2 Frequencies of Local Routes

| Route No. | Route name and section | Number of services | |
|---|---|---|---|
| | | Inbound/East | Outbound/West |
| **30** | **Town Sneaker North Course** | **18** | **–** |
| | *- Loop* | *18* | *–* |
| **31** | **Town Sneaker South Course** | **12** | **–** |
| | *- Loop* | *12* | *–* |
| **32** | **Town Sneaker East Course** | **34** | **–** |
| | *- Loop* | *34* | *–* |
| **33** | **South Circulation Line** | **8** | **–** |
| | *- Loop* | *8* | *–* |
| **34** | **General Administration Liner** | **1** | **1** |
| | *-Matsumoto Sta. – General Administration Office* | *1* | *1* |
| **36** | **Matsumoto / Shimauchi Line** | **13** | **13** |
| | *- Matsumoto Sta. – Lala Matsumoto (Marunouchi Hospital)* | *5 (0)* | *7 (1)* |
| | *- Matsumoto Sta. – Komiya Hall (General Administration Office)* | *7 (1)* | *5 (0)* |
| **37** | **Minami-Matsumoto / Yamagata Line** | **9** | **8** |
| | *- Matsumoto Cinema Lights – I City 21 (Rinku Industrial Complex)* | *5 (1)* | *6 (0)* |
| | *- Minami-Matsumoto Sta. – I City 21 (Rinku Industrial Complex)* | *2 (1)* | *1 (1)* |
| **38** | **Azusagawa / Hata Line** | **6** | **5** |
| | *- Hata Sta. – Azusabashi Sta. (Yamato Park)* | *5 (1)* | *3 (0)* |
| | *- Hakkeiyama Community Center – Azusabashi Sta. (Anzu)* | *0 (0)* | *1 (1)* |
| **39** | **Murai / Yamagata Line** | **10** | **12** |
| | *- Matsumoto Medical Center – I City 21 (Yamagata Vil. Office)* | *4 (1)* | *5 (0)* |
| | *- Murai Sta. – I City 21 (Yamagata Vil. Office)* | *3 (2)* | *2 (5)* |
| **40** | **Asahi / Hata Line** | **8** | **8** |
| | *- Azusagawa High School - Nakagumi* | *1* | *2* |
| | *- Hata Sta. – Nakagumi (I City 21, Matsumoto Nursery School)* | *2 (5)* | *3 (3)* |
| **41** | **Minami-Matsumoto / Hirata Line** | **6** | **6** |
| | *- Minami-Matsumoto Sta. – Hirata Sta.* | *6* | *6* |
| **42** | **Hirata / Murai Line** | **5** | **6** |
| | *- Loop* | *2* | *3* |
| | *- Hirata Sta. – Murai Sta. (Matsumoto Medical Center)* | *1 (2)* | *1 (1)* |
| **45** | **Hata Circulation Line** | **3** | **3** |
| | *- [TUE, WED, FRI] Momo Danchi Depot – Kiroro* | *1* | *1* |
| | *- [MON, WED, THU] 12 District Community Center – Kiroro* | *1* | *1* |
| | *- Sanroku Shimashima Hall – Kiroro (Civil Hospital)* | *1* | *1* |
| **46** | **Hoshimi Line** | **4** | **4** |
| | *- Matsumoto Bus Term. N. – Ichinose* | *4* | *4* |
| **47** | **Iriyamabe Line** | **7** | **10** |
| | *- Matsumoto Bus Term. N. – Owago* | *1* | *1* |
| | *- [FRI] Matsumoto Bus Term. N. – Komagoe* | *2* | *3* |
| | *- [MON, WED, THU] Matsumoto Bus Term. N. – Ushitade* | *2* | *3* |
| | *- [TUE] Matsumoto Bus Term. – Obotoke Community Center* | *2* | *3* |
| **48** | **Nakayama Line** | **4** | **6** |
| | *- Matsumoto Bus Term. N. – Niju Horinomatsu* | *2* | *3* |
| | *- [MON, WED, FRI] Matsumoto Bus Term. N. – Rose Garden* | *2* | *3* |
| **49** | **Asama / Omura Line** | **4** | **4** |
| | *- Matsumoto Castle / City Office – Harahashi W.* | *4* | *4* |

Notes: Only services that operate in the city planning area are summarized.

# Appendix 3 ECO grid distribution in $T = 60$

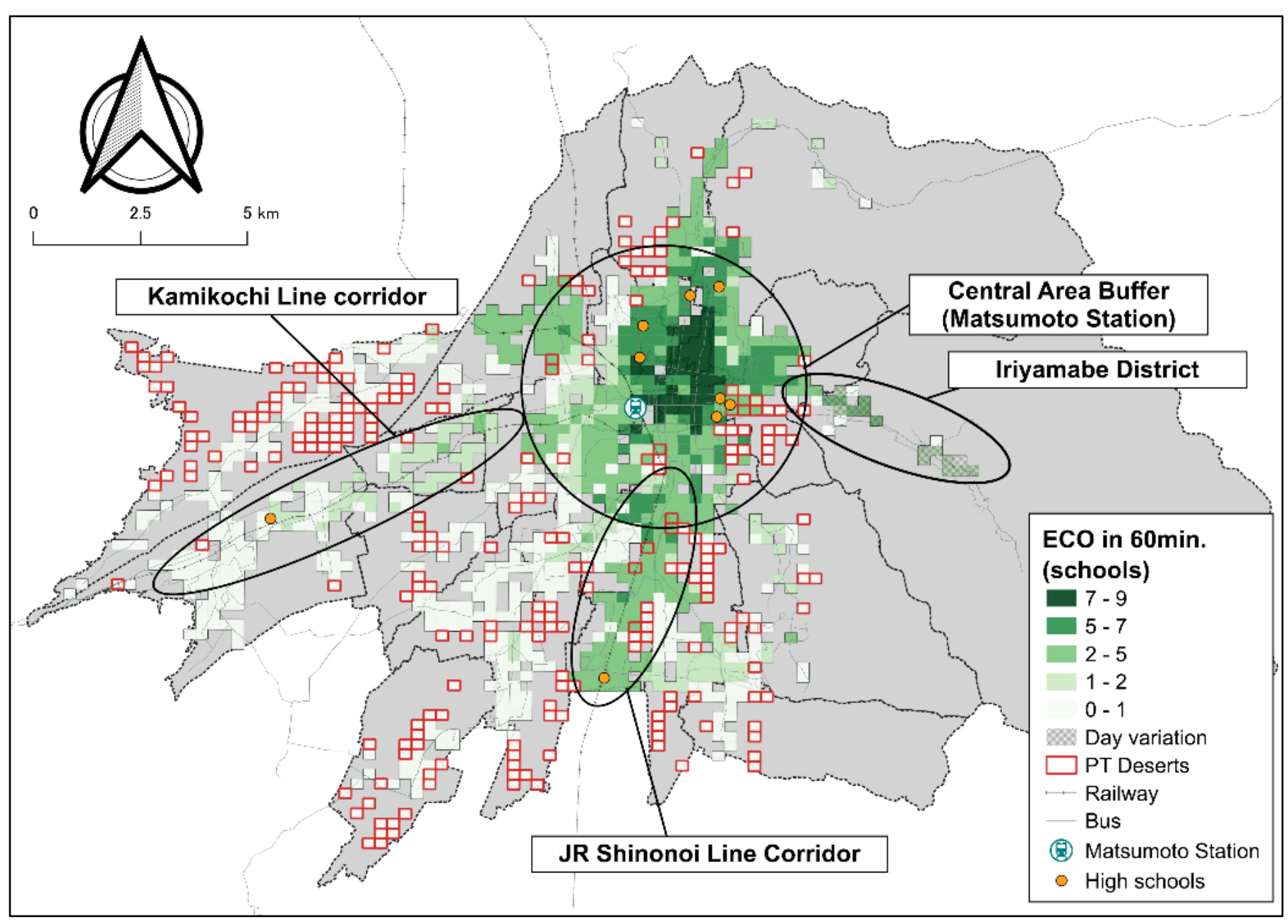

# Appendix 4 ECO grid distribution in $T = 120$

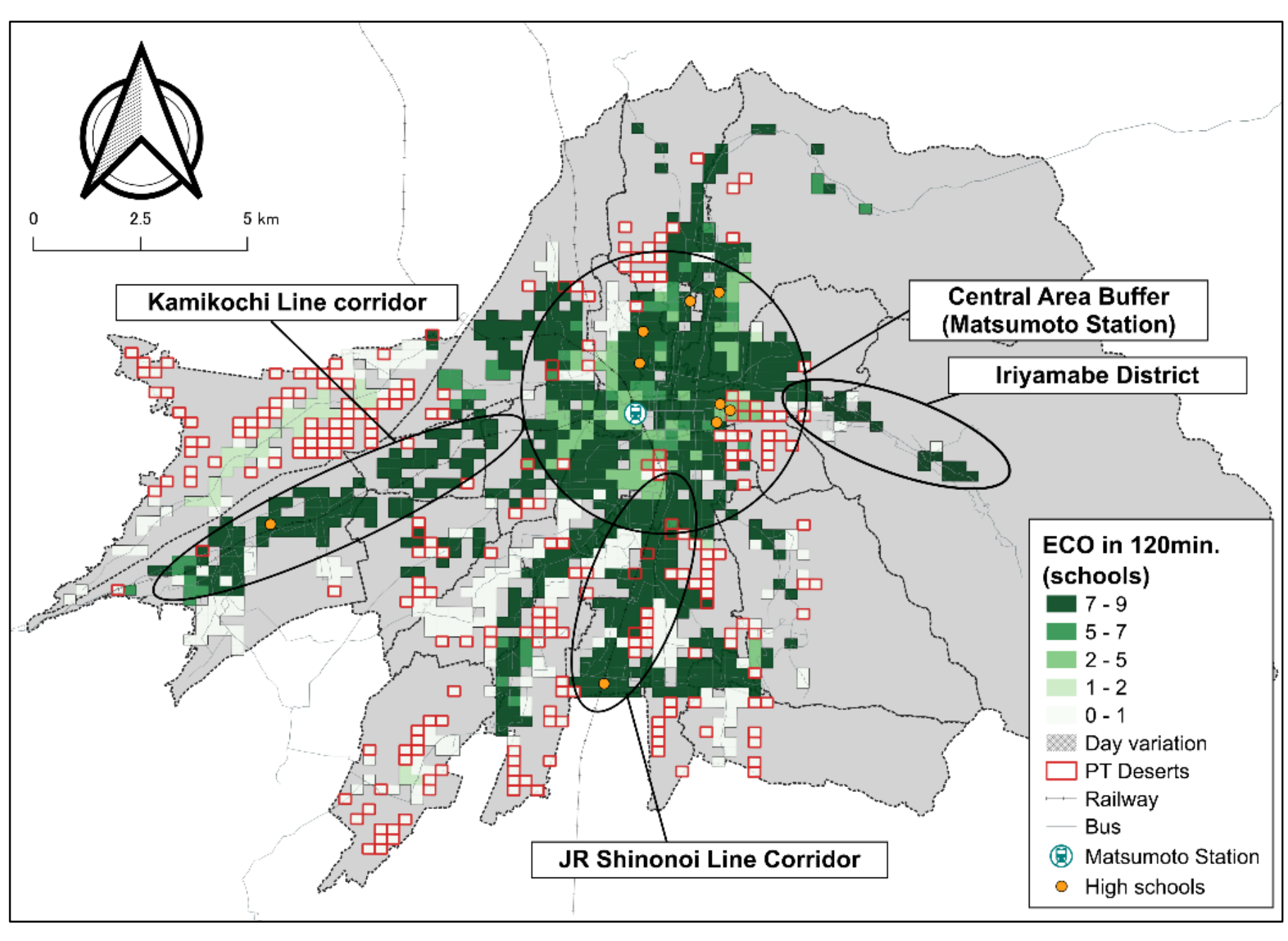

# Appendix 5 Population percentile of ECO (50th)

| District | U15 Population | Thresholds | | | Area |
|---|---|---|---|---|---|
| | | $T = 60$ | $T = 90$ | $T = 120$ | |
| Azusagawa | 1,815 | 0.000 | 0.000 | 0.000 | Azusagawa |
| Chuo | 310 | 6.246 | 6.842 | 7.000 | Central |
| Daiichi | 104 | 2.450 | 2.967 | 3.000 | Central |
| Daini | 233 | 8.134 | 8.827 | 9.000 | Central |
| Daisan | 474 | 6.369 | 8.464 | 8.917 | Central |
| Tobu | 302 | 7.441 | 8.601 | 8.953 | Central |
| Johoku | 1,027 | 5.854 | 7.782 | 8.735 | Central North |
| Joto | 451 | 7.403 | 8.499 | 8.874 | Central North |
| Shiraita | 750 | 4.366 | 7.658 | 7.999 | Central North |
| Yasuhara | 582 | 6.413 | 8.195 | 8.731 | Central North |
| Kamada | 2,731 | 1.785 | 6.319 | 8.581 | Central South |
| Shonai | 1,832 | 4.210 | 7.958 | 8.809 | Central South |
| Shonan | 612 | 4.271 | 8.262 | 8.921 | Central South |
| Tagawa | 485 | 2.665 | 7.669 | 9.000 | Central South |
| Hata | 2,107 | 1.000 | 4.400 | 7.550 | Hata |
| Okada | 1,051 | 4.073 | 7.093 | 8.598 | Higashiyama North |
| Hongo | 1,598 | 5.174 | 7.476 | 8.164 | Higashiyama North |
| Iriyamabe | 169 | 1.751 | 5.542 | 7.799 | Higashiyama Middle |
| Satoyamabe | 1,534 | 4.168 | 7.268 | 8.468 | Higashiyama Middle |
| Nakayama | 293 | 0.161 | 1.918 | 3.320 | Higashiyama South |
| Uchida | 294 | 0.000 | 0.000 | 0.000 | Higashiyama South |
| Niimura | 350 | 1.490 | 6.382 | 8.791 | Kasai North |
| Shimadachi | 853 | 1.153 | 6.388 | 8.719 | Kasai North |
| Shimauchi | 1,801 | 3.247 | 7.503 | 8.814 | Kasai North |
| Imai | 386 | 0.737 | 1.000 | 1.000 | Kasai South |
| Kambayashi | 584 | 0.000 | 0.700 | 1.000 | Kasai South |
| Sasaga | 1,401 | 0.412 | 3.761 | 6.821 | Kasai South |
| Wada | 659 | 0.066 | 5.134 | 8.215 | Kasai South |
| Kotobuki | 1,897 | 1.000 | 4.445 | 7.391 | South |
| Kotobukidai | 276 | 1.299 | 5.717 | 8.255 | South |
| Matsubara | 304 | 1.418 | 6.013 | 8.547 | South |
| Yoshikawa | 2,500 | 3.759 | 7.886 | 8.853 | South |

Notes: All the indices were calculated based on the 25th percentile of the travel-time distribution.

# Appendix 6 Population percentile of ECO (75th)

| District | U15 Population | Thresholds | | | Area |
|---|---|---|---|---|---|
| | | $T = 60$ | $T = 90$ | $T = 120$ | |
| Azusagawa | 1,815 | 0.000 | 0.000 | 0.000 | Azusagawa |
| Chuo | 310 | 6.072 | 6.824 | 7.000 | Central |
| Daiichi | 104 | 2.450 | 2.967 | 3.000 | Central |
| Daini | 233 | 3.537 | 4.676 | 5.000 | Central |
| Daisan | 474 | 6.061 | 7.616 | 7.966 | Central |
| Tobu | 302 | 7.138 | 8.591 | 8.951 | Central |
| Johoku | 1,027 | 2.551 | 4.012 | 4.734 | Central North |
| Joto | 451 | 7.178 | 8.459 | 8.844 | Central North |
| Shiraita | 750 | 1.159 | 2.532 | 2.910 | Central North |
| Yasuhara | 582 | 3.800 | 5.255 | 5.791 | Central North |
| Kamada | 2,731 | 0.966 | 2.861 | 3.000 | Central South |
| Shonai | 1,832 | 3.264 | 7.238 | 8.479 | Central South |
| Shonan | 612 | 0.712 | 5.813 | 8.535 | Central South |
| Tagawa | 485 | 2.346 | 3.000 | 3.000 | Central South |
| Hata | 2,107 | 1.000 | 3.884 | 7.163 | Hata |
| Okada | 1,051 | 3.249 | 6.173 | 7.936 | Higashiyama North |
| Hongo | 1,598 | 2.415 | 4.135 | 5.638 | Higashiyama North |
| Iriyamabe | 169 | 0.908 | 5.007 | 7.685 | Higashiyama Middle |
| Satoyamabe | 1,534 | 2.502 | 3.000 | 3.000 | Higashiyama Middle |
| Nakayama | 293 | 0.000 | 0.000 | 0.000 | Higashiyama South |
| Uchida | 294 | 0.000 | 0.000 | 0.000 | Higashiyama South |
| Niimura | 350 | 1.092 | 6.003 | 8.666 | Kasai North |
| Shimadachi | 853 | 0.059 | 2.774 | 3.000 | Kasai North |
| Shimauchi | 1,801 | 1.928 | 5.637 | 6.532 | Kasai North |
| Imai | 386 | 0.000 | 0.000 | 0.000 | Kasai South |
| Kambayashi | 584 | 0.000 | 0.376 | 0.782 | Kasai South |
| Sasaga | 1,401 | 0.000 | 0.879 | 1.000 | Kasai South |
| Wada | 659 | 0.000 | 3.906 | 7.430 | Kasai South |
| Kotobuki | 1,897 | 0.000 | 0.000 | 0.000 | South |
| Kotobukidai | 276 | 1.299 | 5.717 | 8.255 | South |
| Matsubara | 304 | 1.126 | 5.555 | 8.331 | South |
| Yoshikawa | 2,500 | 0.599 | 1.000 | 1.000 | South |

Notes: All the indices were calculated based on the 25th percentile of the travel-time distribution.

**Acknowledgements**

This research was supported by JSPS KAKENHI Grant Number 23K26220.
We also thank the Matsumoto City Planning Department for providing GIS data.

# Declarations

The authors declare no conflicts of interest regarding this manuscript. During the preparation of the manuscript, the authors used DeepL and ChatGPT in order to refine English writing. After using the tool, the authors reviewed and edited the content as needed and take full responsibility for the contents of the published article.